\documentclass[acmlarge]{acmart}

\usepackage{epsfig,endnotes}
\usepackage{amsmath,amssymb}
\usepackage{url}
\usepackage{multirow}
\usepackage{color}
\usepackage{diagbox}
\usepackage{subfigure}
\usepackage{balance}
\usepackage{hyperref}
\usepackage[hyphenbreaks]{breakurl}
\usepackage{algorithm}
\usepackage{algorithm,algpseudocode}
\usepackage{tablefootnote}
\usepackage{booktabs}
\usepackage{graphicx}

\newcommand{\para}[1]{{\vspace{3pt} \noindent \emph{\bf #1} \hspace{6pt}}}

\newcommand\fixme[1]{{\bf \color{red} #1}}

\newenvironment{packed_itemize}{
\begin{list}{\labelitemi}{\leftmargin=1.em}
  \setlength{\itemsep}{2pt}
  \setlength{\parskip}{0pt}
  \setlength{\parsep}{0pt}
  \setlength{\headsep}{0pt}
  \setlength{\topskip}{0pt}
  \setlength{\topmargin}{0pt}
  \setlength{\topsep}{0pt}
  \setlength{\partopsep}{0pt}
}{\end{list}}

\newfont{\mycrnotice}{ptmr8t at 7pt}
\newfont{\myconfname}{ptmri8t at 7pt}
\begin{document}

%\title{Bracelets of Silence: Personalized Audio Privacy using Wearable Jammers}

\title{Understanding the Effectiveness of Ultrasonic Microphone Jammer}

\author{Yuxin Chen}
\authornote{Both authors contributed equally to this research.}
\email{yxchen@cs.uchicago.edu}
\author{Huiying Li}
\authornotemark[1]
\email{huiyingli@cs.uchicago.edu}
\affiliation{%
  \institution{University of Chicago}
}

\author{Steven Nagels}
\email{stevennagels@cs.uchicago.edu}
\affiliation{%
  \institution{University of Chicago}
}

\author{Zhijing Li}
\email{zhijing@cs.ucsb.edu}
\affiliation{%
  \institution{UC Santa Barbara}
}

\author{Pedro Lopes}
\email{pedrolopes@cs.uchicago.edu}
\affiliation{%
  \institution{University of Chicago}
}

\author{Ben Y. Zhao}
\email{ravenben@cs.uchicago.edu}
\affiliation{%
  \institution{University of Chicago}
}

\author{Haitao Zheng}
\email{htzheng@cs.uchicago.edu}
\affiliation{%
  \institution{University of Chicago}
}

\begin{abstract}
Recent works have explained the principle of using ultrasonic transmissions
to jam nearby microphones. These signals are inaudible to nearby users, but
leverage ``hardware nonlinearity'' to induce a jamming signal inside
microphones that disrupts voice recordings. This has great implications on
audio privacy protection.

In this work, we gain a deeper understanding on the effectiveness of
ultrasonic jammer under {\em practical scenarios}, with the goal of disabling both visible and hidden microphones in the surrounding area. We first experiment with existing jammer designs (both commercial products and that proposed by recent papers), and find that they all offer limited angular coverage, and can only target microphones in a particular direction. We overcome this limitation by building a circular transducer array as a wearable bracelet. It emits ultrasonic signals simultaneously from many directions, targeting surrounding microphones without needing to point at any. More importantly, as the bracelet moves with the wearer, its motion increases jamming coverage and diminishes blind spots (the fundamental problem facing any transducer array). We evaluate the jammer bracelet under practical scenarios, confirming that it can effectively disrupt visible and hidden microphones in the surrounding areas, preventing recognition of recorded speech. We also identify limitations and areas for improvement.

% Home digital assistants and voice-enabled smart devices are ubiquitous
% today. These devices have always-on microphones, and are privy to some of our
% most private conversations. Unfortunately, recent events show that whether
% it's compromise by attackers or bugs in software, these devices can leak
% audio to third parties without user knowledge.

% How can users be assured of their voice privacy in this age of voice-enabled
% devices?  In this paper, we design and validate a wearable microphone jamming bracelet, which
% when activated by the wearer, transmits carefully designed audio signals in
% the ultrasonic frequency range. These signals are inaudible to nearby users,
% but leverage ``nonlinearity'' properties of commodity microphones to induce a
% jamming signal inside microphones that disables voice recordings. Using both
% controlled experiments and natural user experiments, we show that our jammer
% prototype can be effective against known and unknown microphones near the
% user. It offers an initial blueprint towards the development of a low-cost
% and ubiquitous microphone jammer.

\end{abstract}




\maketitle

%\category{}{}

%\begin{figure}[t]
 %   \centering%
%	\includegraphics[width=1\textwidth]{figs/prototypes/wearable/fig1.pdf}
%%\hspace{0.25in}
%%	\includegraphics[width=0.45\textwidth]{figs/walkthrough_2-lessLabels.pdf}
%%\vspace{-0.05in}
% \caption{A user touches our prototype wearable bracelet to enter private
 %  mode, which emits ultrasonic signals (inaudible to
  % the users) to jam any microphones several meters from the
  % user in any direction.}
  %\label{fig:system}
%\end{figure}

\section{Introduction}
\label{sec:intro}
Despite the initial excitement around voice-based smart devices for the
home and office, consumers are becoming increasingly nervous with the fact that
these smart devices are, by default, {\em always} listening,
recording, and possibly saving sensitive personal information they
hear~\cite{nytime-alexa,csmonitor2017,wired2016recordvoice,google-24listening}. Take
home digital assistants as an example.
From the outside, they appear to only respond to designated wake-up
words ({\em e.g.\/}``Alexa'' and ``Hey Google''). However, their
implementation requires them to listen continuously to detect these
wake-up words.  It has been shown that these devices can monitor and record
all voices, sounds and conversations in real time, either
maliciously~\cite{symantec-voice-threat}, by
misconfiguration~\cite{google-24listening}, or after compromise by
attackers~\cite{SongCoRR17}. Leaked audio data can be processed to
extract confidential
information~\cite{symantec-voice-threat,Chung2018life,Chung2017}, track user
activity~\cite{Awaki:2016ubicomp}, count
speakers~\cite{Xu:2013ubicomp},  or even extract handwriting
content~\cite{Yu:2016ubicomp}.  These negative implications on users'
security and privacy are significant and unacceptable.

Clearly, it is important to build tools that  can protect users against the potential compromise or
misuse of microphones in the age of voice-enabled smart-devices.
Recent work along this line~\cite{backdoor} shows that ultrasonic microphone
jammers can emit an ultrasonic wave that prevents commodity microphones
from recording human speech. While these ultrasonic signals are
imperceptible to human ears, they leak into the audible spectrum after
being captured by commodity microphones, producing a jamming signal
inside the microphone circuit to disrupt voice recordings. The
leakage is caused by an inherent, nonlinear property of microphone
hardware. Not only have researchers built low-cost prototypes using off-the-shelf ultrasonic
speakers~\cite{backdoor}, but also ultrasonic jammers are currently even commercially available to the public.

In this work, we seek a deeper understanding of this approach by
studying the effectiveness of ultrasonic microphone jammers under
practical scenarios. Current studies~\cite{backdoor,roynsdi18,dolphin}
focus on disabling a known microphone device by pointing the jammer at
it. In contrast, our work considers broader and more complex everyday
scenarios. We explore (1) jamming both visible and hidden microphones in
an area, (2) strategies to minimize blind spots in coverage of
current ultrasonic devices, and (3) jamming under realistic
scenarios where either the human speaker or the microphones are moving.

Our work is organized into three phases.

\begin{figure}[h!]
 \centering
\mbox{
\subfigure[Our wearable ultrasonic jammer moves with the user]{
   \includegraphics[width=0.465\textwidth]{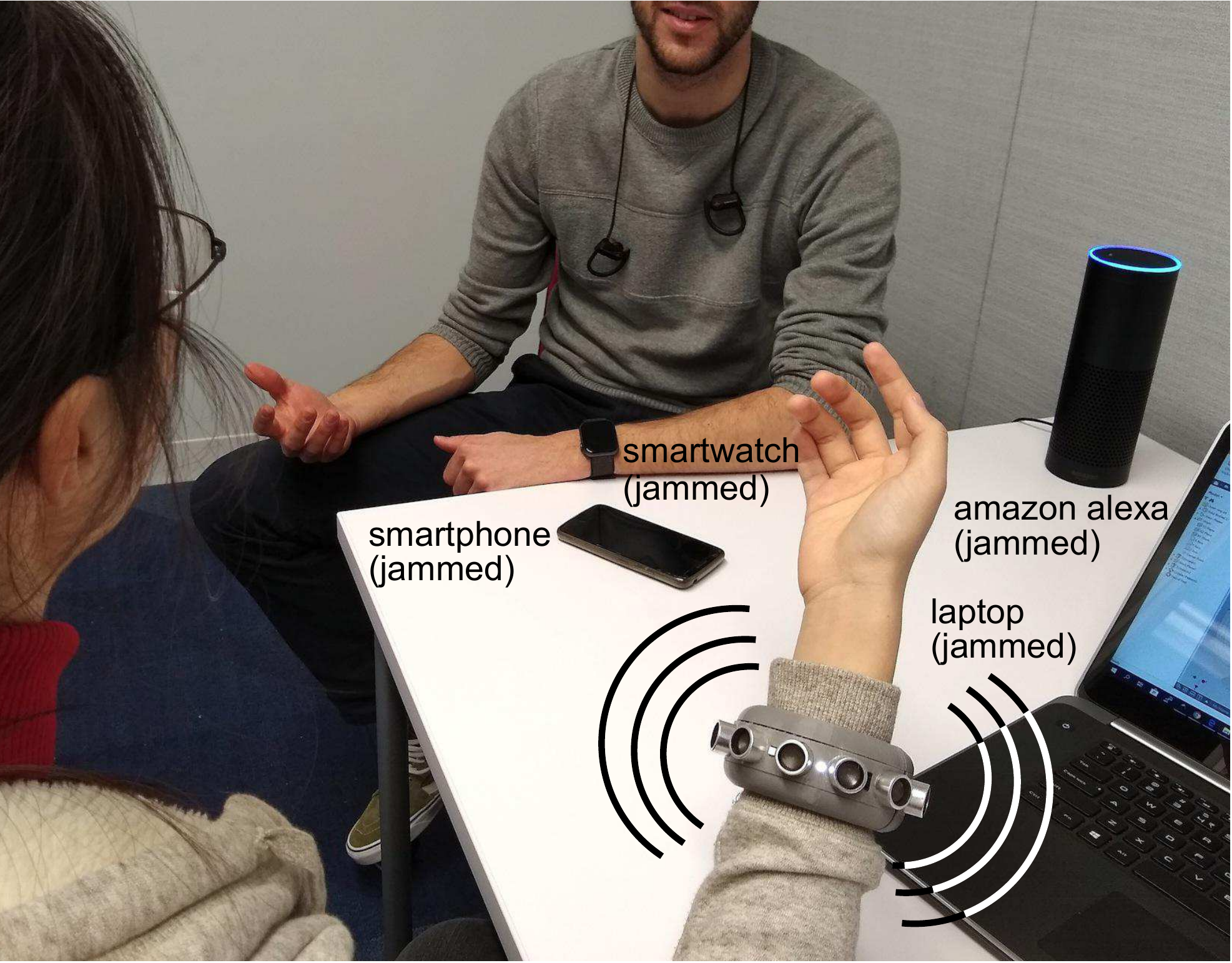}
    \label{fig:placement*}
   }
   %\hspace{1mm}
 \subfigure[Commercial ultrasonic jammer from i4, \$799]{
   \includegraphics[width=0.235\textwidth]{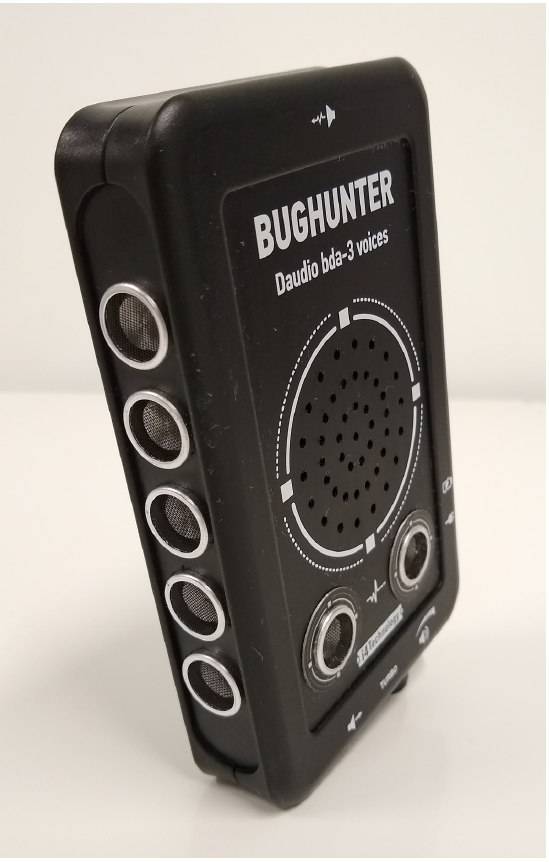}
    \label{fig:i4_jammer*}
   }
   \hspace{1mm}
   \subfigure[\textit{Backdoor} 3$\times$3 jammer~\cite{backdoor} using ultrasonic transducers]{
   \includegraphics[width=0.235\textwidth]{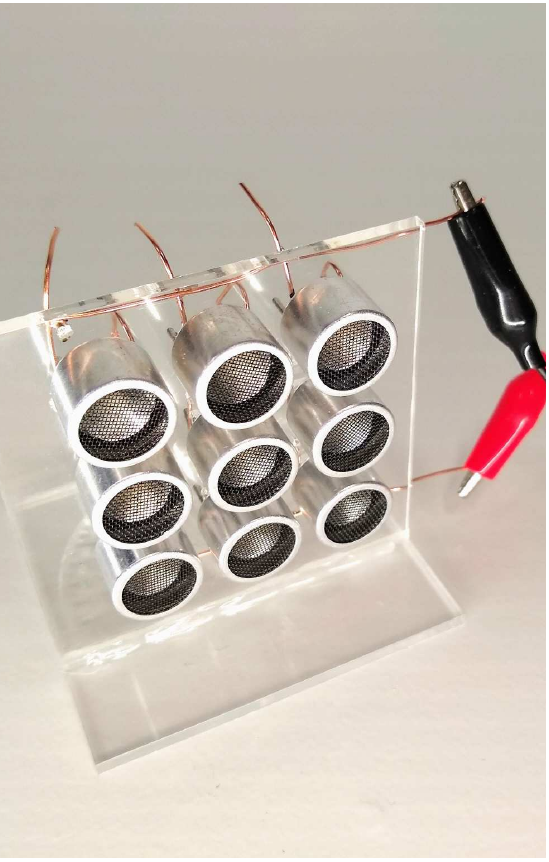}
    \label{fig:lab_made*}
   }
 }
 \vspace{-0.15in}
\caption{We demonstrate that ultrasound jammers can be more effective if made
  wearable (a), instead of the current approach, which is to use stationary
  emitters (b,c). Our prototype is a bracelet that jams surrounding
  microphones using ultrasound, leveraging the known effect of the
  microphone's non linearity~\cite{backdoor,roynsdi18,dolphin}. We designed
  and validated the effectiveness of this wearable jamming bracelet in
  comparison to an (b) existing commercial and (c) state of the art ultrasound
  jammers. We found that our approach offers omni-directional jamming, increases
coverage, removes undesired blind spots, and requires less
  power than commercial jammers.} \label{fig:intro}
\end{figure}

First, we experiment with two of today's ultrasonic jammer platforms,
including a commercial product~(Figure~\ref{fig:i4_jammer*}) and a prototype suggested by a recent
research paper~\cite{backdoor}~(Figure~\ref{fig:lab_made*}).  We test both
jammers, and find that they offer only directional jamming with limited angular
coverage and produce blind spots within the covered directions.  This is caused by the inherent directionality of commodity ultrasonic
transducers and the use of transducer arrays. Since the ``speakers''
used in these devices operate beyond the audible range, they are
denoted with ``ultrasonic transducer,'' and we will refer to these simply as ``transducers.''

In the second phase, we expand angular coverage by placing multiple
transducers on a wearable bracelet, which simultaneously emits ultrasound
in many directions. Thus the bracelet jammer mimics an {\em
  omni-directional} jammer using inherently {\em directional} off-the-shelf
ultrasonic transducers.  More importantly, as the wearable jammer moves with the
wearer, normal motion by the user effectively increases coverage and
dramatically reduces coverage blind spots (the fundamental problem facing any
speaker array).  We implemented our design into a self-contained, wearable,
jamming bracelet that we depict in Figure~\ref{fig:placement*} and
later in Figure~\ref{fig:bracelet}.

%does not require one jammer per room,

%   \centering
%   %\mbox{
%   \includegraphics[width=0.41\textwidth]{figs_new/wearable2.0/Conversation-with-labels.pdf}
%    %figs/user-gesture.pdf}
% 	\label{fig:opening}
%     %}
%    %\vspace{-0.15in}
% 	\caption{We demonstrate that ultrasound jammers can be more effective if made wearable, instead of the current approach is to use stationary emitters. Our prototype is a wearable bracelet that jams surrounding microphones via ultrasound emission, leveraging the known effect of the microphone's non linearity~\cite{backdoor,roynsdi18,dolphin}. We designed and validated the effectiveness of this wearable jamming bracelet in comparison to both existing commercial  and state of the art ultrasound jammers. We found that our approach inherently increases room coverage (since the user moves alongside with the jammer), does not require one jammer per room, decreases blind spots (since the user's naturally-occurring hand gestures increase coverage) and requires less power than existing jammers.}
% \end{figure}

Finally, we study the effectiveness of this new wearable ultrasonic jammer in
natural settings that have not been considered by prior work, including
scenarios with multiple microphone devices, hidden (or covered) microphone
devices, multiple users engaged in conversations, and even users in motion while talking.  We use these
experiments to validate the effectiveness of the our jammer design, and
to identify its limitations and areas for improvement.

\para{Our Contributions.}
\begin{packed_itemize}

\item Understanding and expanding the angular coverage of ultrasonic microphone jammers.
\item Increasing jamming coverage through the design of a wearable jammer, which in addition leverages
  the user's gestures to further reduce blind spots.
\item Systematic evaluation in life-like scenarios to validate effectiveness
  and identify practical limitations of ultrasonic jammers.
\end{packed_itemize}

\section{Background and Related Work}
\label{sec:related}

% Our work builds on top of \fixme{state exactly what sub areas of ubicomp/CS this builds on}. Also, we discuss alternative methods of jamming microphone recordings and their inherent limitations. Lastly, we introduce the underlying ultrasonic jamming principle that our device leverages.

% Recent work has shown home digital assistants to be vulnerable to many
% malicious attacks, ranging from bypassing
% permissions~\cite{Diao2014YourVA}, voice
% impersonation~\cite{chenicdcs17,lei2017arxv}, to multiple methods of
% issuing inaudible hidden commands~\cite{carlini2016hidden,
%   Carlini2018,kasmi2015,roynsdi18,SongCoRR17,Vaidya2015CocaineNE,dolphin}. In
% this context, there are some strong defenses possible.

As background, we describe the underlying principle behind ultrasonic
microphone jammers and summarize prior work in this area. We also briefly discuss
prior work that leverages ultrasonic signals for both sensing and
communication to contextualize the usage of ultrasound in HCI.

\subsection{Principles of ``Silent'' Microphone Jamming using Ultrasonic Signals}

Recent work demonstrated the feasibility of using ultrasonic transducers
to disable nearby microphones. Such jamming is ``silent'' since ultrasound is
inaudible\footnote{Ultrasound is sound waves of
  frequencies above the upper bound of human hearing (20kHz).} to most
humans. Such jamming is possible because ultrasonic signals, after
being captured by commodity microphones (MEMS microphones), leak into the
audible spectrum and produce a jamming signal {\em inside} the
microphone circuit. This leakage is caused by {\em hardware non-linearity},
an inherent property of commodity microphone
devices~\cite{abuelma2003analysis:1}. This leakage adds so much audible noise
on the microphone circuitry that it effectively renders voice recordings
unusable.

% \para{Properties of Ultrasound.}
% Ultrasound is sound waves that have frequencies above the upper bound of human hearing.
% The frequency range of human hearing is generally considered as [20, 20000] Hz and
% ultrasound is considered as the sound waves of frequencies higher than 20 kilohertz
% which is inaudible for human. Ultrasound is used in many different fields, such
% as diagnostic medical sonography~\cite{diagnosticmedicalsonography}, non-destructive
% testing (NDT)~\cite{ultrasoundforNDT}, cleaning, mixing as well as communication
% ~\cite{backdoor}.

% In physics, attenuation is the gradual loss of flux intensity through a medium.
% Absorption of sound through the atmosphere causes the attenuation of sound waves.
% Attenuation coefficient is a metrics that describes the extent to which the
% radiant flux of a sound wave attenuates as it passes through a specific material.
% It is dependent on the pressure, relative humidity and frequency of the sound
% wave.

\para{Non-linearity in Microphone Hardware.}  Linearity in microphones refers
to its ability to generate an electrical output proportional to the amplitude
of the sound input. While electronic components such as amplifiers are
carefully designed to be linear over as wide a frequency range as possible,
linear recording devices do not exist in practice. Any device, such as a
microphone, exhibits non-linearity in some frequency bands. This
non-linearity in microphones was originally discovered by musicians and
leveraged for sound synthesis~\cite{IMDoriginal}. Only more recently has it
generated serious impact on the mobile and security communities,
given the pervasiveness of microphones in digital voice assistants and
smartphones~\cite{backdoor,dolphin,SongCoRR17,roynsdi18}.

% \begin{wrapfigure}{r}{0.42\textwidth}
%     \centering
%     \vspace{-0.1in}
 %   \includegraphics[width=0.4\textwidth]{figs/signalflow.pdf}
  %  %\vspace{-0.1in}
 %\caption{\small Signal flow of a microphone.}
 % \label{fig:signalflow}
%\end{wrapfigure}

%As shown in Figure~\ref{fig:signalflow},

A commodity microphone consists of four components: a transducer, an amplifier, a low-pass filter, and an analog-to-digital converter
(ADC). The low-pass filter has a cut-off frequency of 20KHz (human audible
range is $[$20Hz,20kHz$]$) to support the ADC. One can
represent the microphone input signal $S_{in}$ and output signal $S_{out}$ as
follows:
%\begin{align}
% S_{out}(t) = \sum_{i=1}^{\infty}{A_i S_{in}^i} = A_1 S_{in} + A_2 S_{in}^2 + A_3 S_{in}^3 + ...
%\end{align}%
%where $A_i$ is the gain for  $S_{in}^i$ and is frequency-dependent.%
%When $S_{in}$ is in the audible frequency, the 2nd and higher order
%terms are extremely weak and will be ignored, while for ultrasonic signals, the 3rd and higher terms
%will be ignored. Thus we have
%One can model the microphone output $S_{out}$ to any
%ultrasonic input $S_{in}$ by  $S_{out} = A_1 S_{in} + A_2 S_{in}^2$,
\begin{align}\vspace{-0.08in}
 S_{out} = \sum_{i=1}^{\infty}{A_i S_{in}^i} = A_1 S_{in} + A_2 S_{in}^2+A_3
 S_{in}^3 + ...
\vspace{-0.06in}
\end{align}
where the 2nd term $A_2 S_{in}^2$ and the subsequent terms reflect the non-linear behavior of the
microphone hardware.

%\begin{figure}[t]
 %    \centering
  %   \vspace{-0.1in}
   % \includegraphics[width=0.4\textwidth]{figs/jammerPipeline.pdf}
   %\vspace{-0.05in}
 %\caption{\small Pipeline of Our Initial Microphone Jammer Design}
 % \label{fig:jammerPipeline}
%\end{figure}

% Figure~\ref{fig:jammerPipeline} shows the pipeline of our stationary
% microphone jammer design, which includes four components: a 5V power supply,
% an ultrasonic signal generator, a class-D audio amplifier (PAM8403) and an
% ultrasonic transducer (NU25C16T-1, 25kHz). We chose these commodity
% hardware units for their high availability and low cost.

\begin{figure}[t]
  \centering
  \mbox{
\subfigure[Ultrasonic jamming signal]{
   \includegraphics[width=0.32\textwidth]{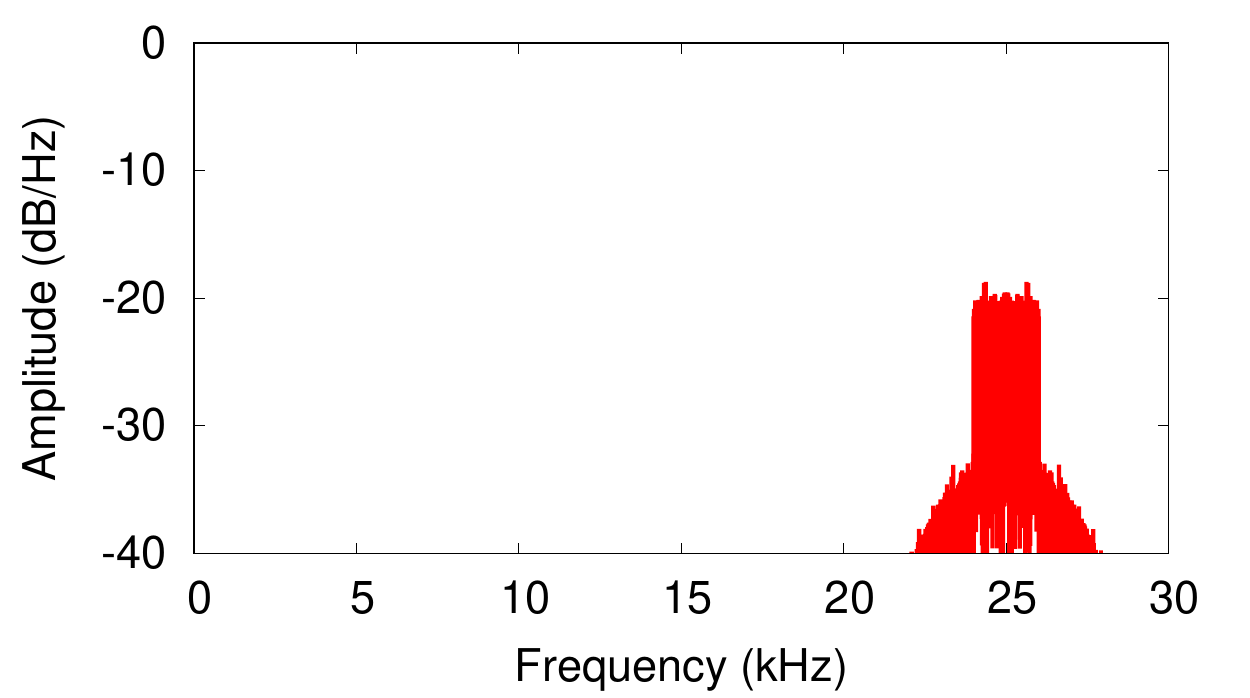}
    \label{fig:ampn}
}
 \subfigure[Microphone recording of (a)]{
   \includegraphics[width=0.32\textwidth]{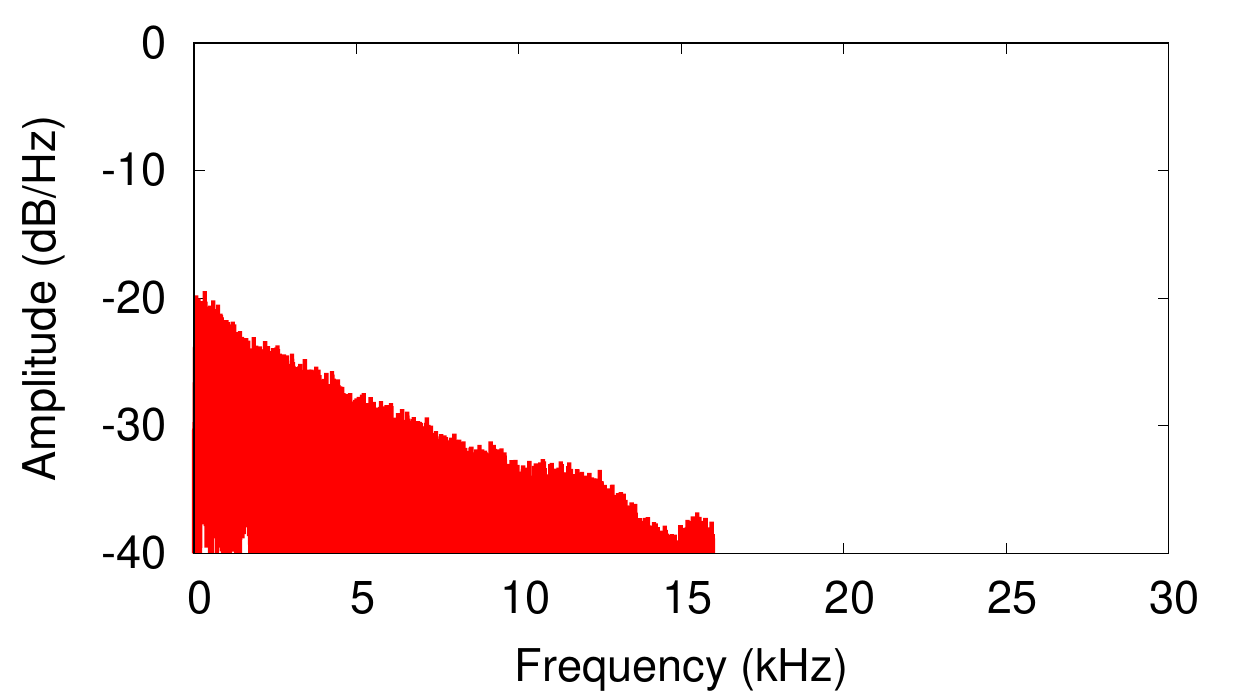}
   \label{fig:ampnr}
   }
\subfigure[Recording of (a) + human speech]{
   \includegraphics[width=0.32\textwidth]{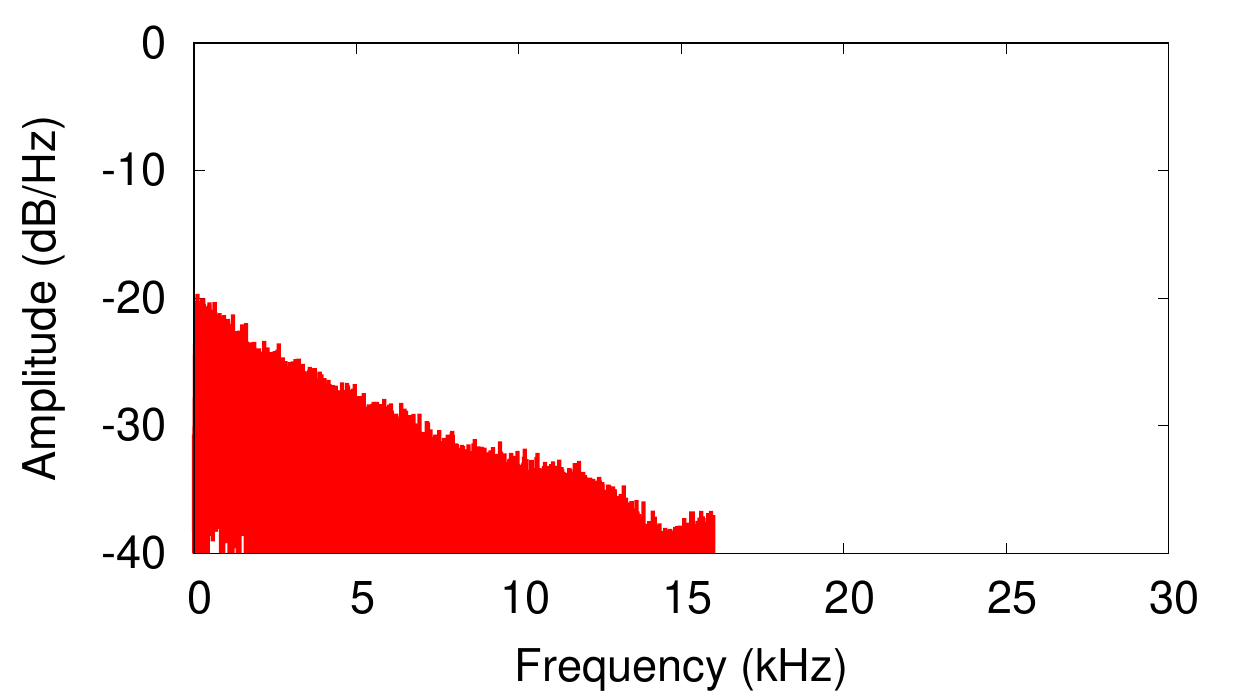}
  \label{fig:ampnrv}
}
}
\vspace{-0.1in}
 \caption{Samples of (a) source ultrasonic jamming signals, (b) after
   being
   captured by the microphone; (c) after being captured by the
   microphone together with human speech.  }
 \label{fig:lbpnam}
\end{figure}

The process of ultrasonic jamming is simple: the signal generator
produces a carefully crafted jamming signal in the ultrasonic band, passes it
to the amplifier and then to the ultrasonic transducer. When captured by nearby
microphones, the jamming signal leaks into the audible band, and distorts any
recordings, particularly those of human voices. Figure~\ref{fig:lbpnam} shows
example traces of the source ultrasonic jamming signal (as amplitude
modulated white noise) and as it is captured
by the microphone both without and with the presence of human speech.  We see
that ultrasonic signals (centered around 25kHz) produce microphone recordings
that cover up signals of human speech.

% When the linear term takes a sinusoidal input signals and
% outputs a sinusoidal signal with the same frequency of input.
% The quadratic term, however, produces harmonics and cross-products.
% In particular, suppose the input signal is sum of two tones with frequencies
% $f_1$ and $f_2$, the input signal can be expressed as
%\begin{align}
 % S_{in}(t) = cos(2 \pi f_1 t) + cos(2 \pi f_2 t)
%\end{align}
%then the output of the quadratic term is
%\begin{equation*}
 % \begin{aligned}
  %%  A_2 S_{in}^2(t) = A_2 +
    %\frac{A_2}{2} ( cos(2 \pi (2 f_1) t) + cos(2 \pi (2 f_2) t) ) + \\
    %A_2 (cos(2 \pi (f_1 + f_2) t) + cos(2 \pi (f_1 - f_2) t))
  %\end{aligned}
%\end{equation*}%
%which contains harmonic frequencies $2 f_1$, $2 f_2$ and cross products
%$f_1 \pm f_2$.

%Figure~\ref{fig:ampn} illustrates the non-linearity of an iPhone X's microphone. Here a
%ultrasonic transducer transmits pink noise at $25kHz$ (cutoff at $22kHz$), while the iPhone X records signals in the $[20Hz,20kHz]$ band owing to the microphone's non-linearity.

%\fixme{Need to find place for the following text}
%\para{Transient or Prevalent Property?}

\subsection{Related Work}

\para{Leveraging Microphone Non-linearity.} Recently, researchers have
leveraged microphone non-linearity as a potential tool for
setting up hidden communication channels, disabling microphones, or as
an adversarial avenue for injecting hidden voice commands.

A series of projects leveraged this property to attack digital voice
assistants~\cite{dolphin,SongCoRR17,roynsdi18}. Here, an adversary can play
(arbitrary) voice commands modulated in the ultrasonic range and leverage the
non-linearity of microphones in home digital assistants ({\em e.g.\/} Amazon
echo) to force the target device to decode them as normal voice
commands. Since the original ultrasonic command is inaudible, the attacker
can successfully issue commands without being detected ({\em i.e.\/}, heard)
by nearby users.

Recent work by Nirupam et. al.~\cite{backdoor} leverages
non-linearity to build inaudible communication among devices and to jam
microphones. The \textit{Backdoor} device utilizes a jamming signal based on either amplitude
modulation (AM) or frequency modulation (FM). \textit{Backdoor} is
tested in a limited set of experiments, ({\em e.g.\/} the jammer pointing to a
single microphone) to validate the design.  In parallel,  there are
already commercial products that use ultrasonics for microphone jamming,
although all of them are bulky (0.38kg--5kg) and pricey (\$799--\$6900)~\cite{i4-jammer, uspy-defeater,advjammer,towera}.

Our work is inspired by these existing works on microphone non-linearity,
particularly \textit{Backdoor}~\cite{backdoor}. However, we dive deeper into this line of
research, to examine the effectiveness of ultrasonic microphone jammers under
practical scenarios, with the goal of disabling both visible and hidden
microphones in the user's surroundings.

%using directional, high-power (2$Watt$) transducer arrays
%to jam a single visible microphone. In contrast, our work focuses on
%developing wearable, low-power (174-446$mWatt$), and omni-directional jammers
%to provide a full private mode even in the presence of multiple {\em hidden}%
%microphone devices.

%The key element is to craft an efficient jamming signal that effectively
%distorts human voices.  For this, we consider the simple amplitude modulation
%(AM) where the amplitude of the carrier wave is in proportion to the
%amplitude of the signal to be transmitted. We also experimented with jamming
%using Frequency Modulation (FM) techniques (both masking and suppression),
%but found no significant advantage over the simpler AM design. Thus we focus
%on AM signals for the remainder of the paper.

\para{Ultrasonic Signals for Device Interactions.}  Researchers in the HCI
community have used signals in ultrasonic
bands~\cite{Aumi:2013ubicomp,Gellerson2007} and near-ultrasonic bands ({\em
  e.g.\/} 18.8kHz)~\cite{Chen:2014ubicomp,Gupta:2012chi} to enable
interaction with/among devices. As an example, Gupta et al., utilize Doppler
shifts in emitted ultrasound to enable a laptop to perform simple gesture
tracking~\cite{Gupta:2012chi}.  A variety of smartphone apps use
ultrasonic signals as beacons to perform device localization and
tracking~\cite{arp2017privacy, ultrasonicads,Fischer:2008}, again based on
their leakage to the audible band.

%Sample citation~\cite{dolphin}

%Finally, hundreds of apps have used ultrasonic signals as beacons to perform (inaudible)
%accurate device localization and tracking~\cite{arp2017privacy, ultrasonicads}.
%Prior work have also shown that current usage of ultrasonic signals could be
%abused to perform various attacks, including stealthy
%deanonymization~\cite{mavroudis2017privacy, ultrasonicsignal}.

%\input{goal}
%\input{jamming}
\begin{figure}[t]
 \centering
%\mbox{
 %\subfigure[Commercial ultrasonic jammer from i4 Technology, \$799]{
  % \includegraphics[width=0.3\textwidth]{figs_new/existing_device_photo/amz.pdf}
   % \label{fig:i4_jammer}
   %}\hspace{0.1in}
   %\subfigure[Lab-made jammer using off-the-shelf ultrasonic speakers]{
   %\includegraphics[width=0.3\textwidth, height=1.3in]{figs_new/existing_device_photo/3x3.pdf}
   % \label{fig:lab_made}
   %}\hspace{0.1in}
\begin{minipage}{0.32\textwidth}
    %\subfigure[Experiment scenario]{
   \includegraphics[width=0.99\textwidth,
   height=1.3in]{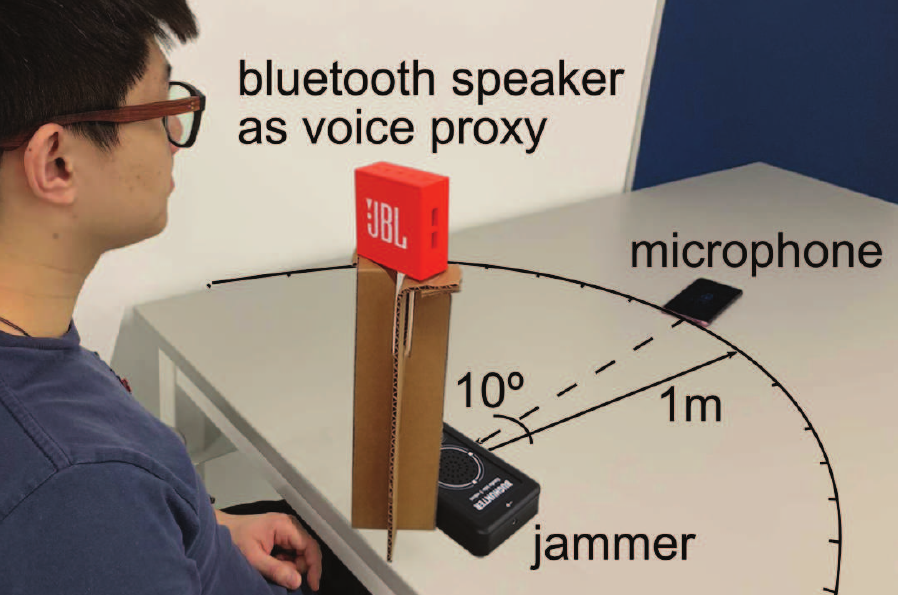}
    \caption{Our evaluation scenario where we vary $\alpha$,  the angular
  separation between the jammer and the microphone device. }
    \label{fig:setup}
   \end{minipage}
 \hfill
\begin{minipage}{0.66\textwidth}
 \subfigure[i4 Tech jammer]{
   \includegraphics[width=0.45\textwidth]{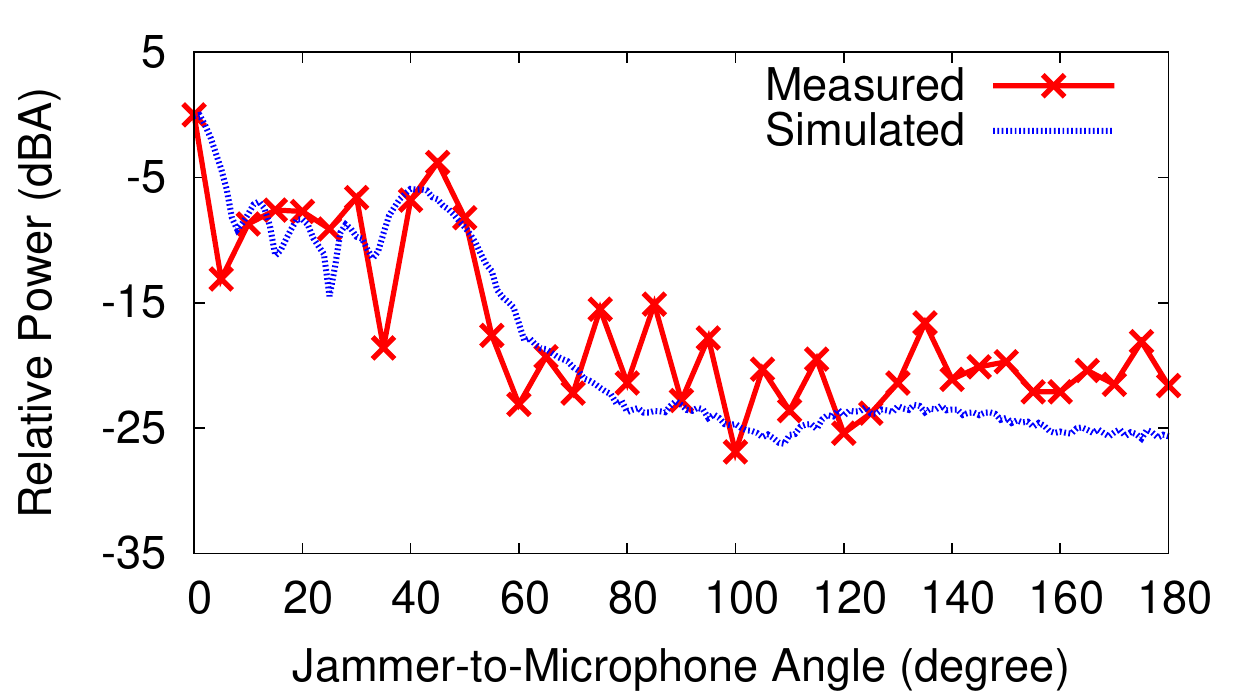}
%figs_new/blind_spots/amz/amz.pdf}
    \label{fig:amz}
}\hfill
  \subfigure[\textit{Backdoor 3x3}]{
    \includegraphics[width=0.45\textwidth]{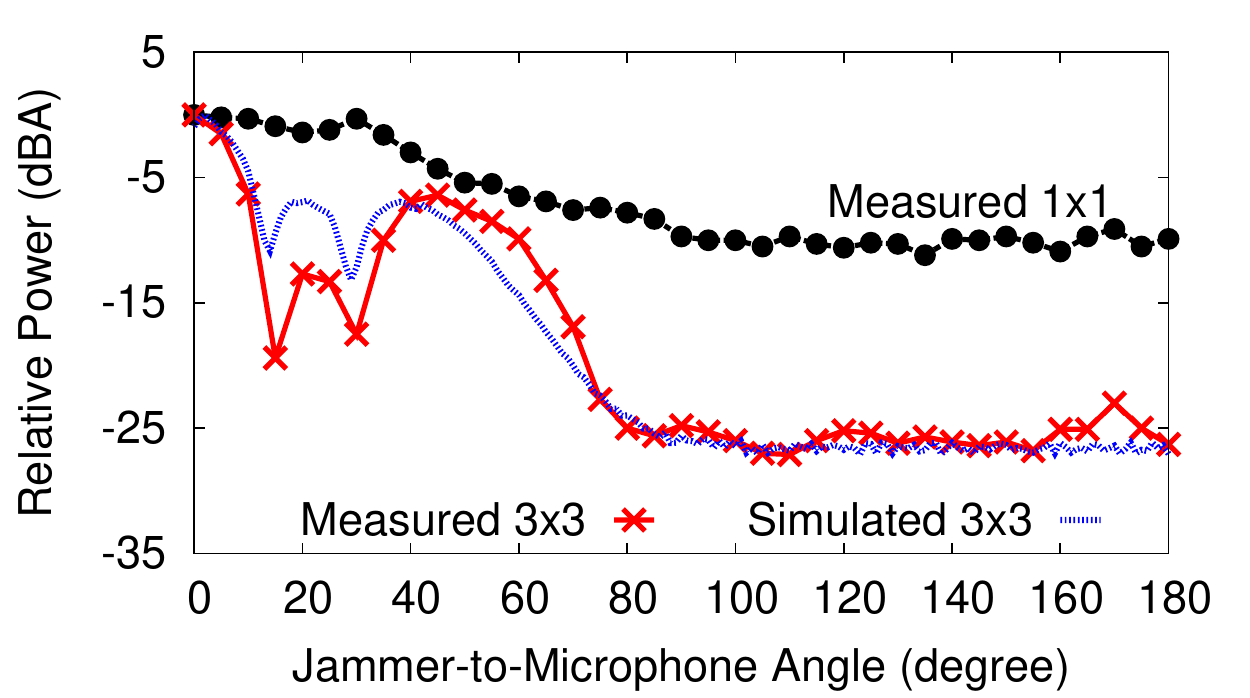}
      %figs_new/blind_spots/3x3/3x3.pdf}
    \label{fig:uiuc}
   }
\vspace{-0.15in}
\caption{Real-world measurements of the jammer's angular
  coverage, in terms of the signal power as  the
  jammer-to-microphone angle $\alpha$ increases from 0$^\circ$ to
  180$^\circ$,  normalized by that of ($\alpha=0^\circ$).   The distance between the jammer  and the microphone is
  kept at 1m.}\label{fig:angle_eval}
\end{minipage}
%\vspace{-0.1in}
%\caption{(a)(b) The two existing ultrasonic jammer devices that we
 % evaluated. (c) Our evaluation scenario where we vary $\alpha$,  the angular
 % separation between the jammer and the microphone device. }
 %\label{fig:experiment_config}
\end{figure}

\section{Evaluation of  Existing Ultrasonic Jammers}
\label{sec:simplevalidation}
We begin our work by evaluating current designs for ultrasonic jamming
devices: (1) a commercial jammer purchased from \textit{Amazon.com} (i4 Technology,
\$750), and (2) the \textit{Backdoor 3x3} jammer that we built using off-the-shelf
ultrasonic transducers following\footnote{We implemented amplitude modulation
  based jamming using a band limited white noise as the ultrasonic source
  signal. The signal bandwidth is 1kHz because it is the operating limit of
  our ultrasonic transducers.} recent work by Nirupam et al.~\cite{backdoor,
  roynsdi18}.  For both devices, we evaluate the jammer's signal coverage
(\S\ref{subsec:signalcoverage}) and its effectiveness in disrupting
microphone recordings~(\S\ref{subsec:speechquality}).

\para{Jammers.} (1) The i4 jammer\footnote{The i4 jammer includes
  a traditional audio jammer (in the audible band) and a ultrasonic
  jammer. We only activate the ultrasonic module.} is shown in
Figure~\ref{fig:i4_jammer*}, and consists of a row of five ultrasonic
transducers on the side and two more on the top.  These transducers operate
at the very low end of ultrasonic frequency (24KHz), and unfortunately even produce
disturbing audible sounds due to signal leakage in the
transducer. This device weighs 380 grams and consumes 4.2W of
power.
(2) The \textit{Backdoor 3x3} jammer, which is depicted in
Figure~\ref{fig:lab_made*}, is an array of nine ultrasonic
transducers. These transducers operate at 25kHz ($\pm 1$Hz) and the sound output
is completely inaudible.  This is not a stand-alone device and its
power supply and circuitry are not integrated.

%They also operate
%at their power limit to maximize jamming power.

% connected in parallel to
% generate a \fixme{2} Watt jamming signal.
% These transducer operate at the 25kHz and produce inaudible
% sounds to human users.

\para{Experimental Setup.} Our evaluation considers a typical scenario in which the ultrasonic jammer is used to jam microphones in the room.  As shown in Figure~\ref{fig:setup},
we placed the jammer on the table and distributed smartphones (serving as microphones)
some distance away.  We performed experiments in four rooms of
varying sizes and furniture arrangements. We found our measurements to
be consistent
across all rooms, thus we present aggregated results.

\vspace{-0.1in}
\subsection{Jamming Coverage and Blind Spots}
\label{subsec:signalcoverage}
Instead of just evaluating the known-effect that distance has on jamming\footnote{We note that the
  physical distance covered by a jammer depends on the ultrasonic transducer'
  power level, the ultrasonic signal frequency, and the volume of the human
  speaker. All of these can vary across scenarios.}, we focus on the angular
coverage of the ultrasonic jammer. Since jammers seek to disrupt microphones
in the surrounding area, angular coverage is a key performance metric; decreasing blind spots is crucial for effective jamming. Our
evaluation used both real-world signal measurements using a sound level meter
at coarse-grained locations, and simulated signal emission maps at
fine-grained locations.

\para{Real-world Measurement.}  We placed a HT-80A sound level meter (which
includes a well-calibrated microphone) 1m away from the
jammer. We moved the sound level meter around the jammer (a 1m radius) to vary the angular
separation between them. Figure~\ref{fig:angle_eval} shows the measurement results of the i4 and
\textit{Backdoor 3x3} jammers in the absence of any human speech. We present
the measured meter power at different degrees of angular separation ($\alpha=0^\circ$ to
180$^\circ$), normalized by that of ($\alpha=0^\circ$).  For the
\textit{Backdoor 3x3} jammer, we also show result for a single ultrasonic transducer.

We made two key observations.

First, {\em both jammers have very limited angular coverage.} As shown, moving away
from a perfect alignment ($\alpha=0^\circ$) results in a drop of the jamming signal strength by 25 to
30dB.  This implies that in order to disrupt microphones in the surrounding
area a user either points directly at the microphones (which is not possible if these are hidden) or each jamming device must raise its transducer's power level by at least 25 to 30dB.  Otherwise
several jammers are necessary to fully cover potential microphones at different
angular positions relative to the user. This lack of angular coverage is
caused by the inherent directionality of commodity ultrasonic transducer, as
shown in Figure~\ref{fig:angle_eval}(b).

{\em Second}, {\em jamming signal power for both jammers shows heavy local
  fluctuations at different jamming angles}.  Even within the angular sector of
[0$^\circ$,40$^\circ$], a subtle angle change of 2$^\circ$ leads to
5-10dB change in jamming power level. This uneven distribution is a
fundamental problem facing transducer arrays, often referred to as the
blind spot problem~\cite{mailloux1982phased}. Mutual coupling of signals
emitted by different transducers creates unevenness in the jammer's emission
pattern, leading to undesired blind spots at certain angular
directions.

\para{Mapping Jamming Power using Propagation Models (Simulation).} To further illustrate
the above two artifacts, we followed the ultrasonic signal propagation
model~\cite{ultrasonicmodel} to generate an ultrasonic signal map for
both jammers. Our simulation used the single transducer's emission pattern, provided
by the manufacturer of the ultrasonic transducer, which we used to replicate the \textit{Backdoor} jammer.  We utilize the same emission pattern to simulate the behavior
of the i4 jammer (since the manufacturer does not provide any information
regarding their transducers).  We marked the jammer location as (0m, 0m)
pointing at (0m, 1m). While the jammer emits white noise signals of 1KHz on
the 25KHz band, we computed the signal power received at each location on the
1m$\times$1m area, normalized by received power at (0m, 0m).

The results plotted in Figure~\ref{fig:heatmap3x3} show the relative jamming
power for the 3$\times$3 jammer. For visual clarity, we omitted the i4 jammer results as they are
very similar. As shown, our simulation confirmed the limited angular coverage (shown as a blue
triangle in the bottom right) and the directions of the blind spots (shown as two blue stripes
in the top left). These simulations are in line with our earlier observations. Note that we also
compared our model-generated signal power values to measured power values in
Figure~\ref{fig:angle_eval}, which we found to be consistent.

\subsection{Speech Recognition under Jamming}
\label{subsec:speechquality}
For an end-to-end evaluation of jamming effectiveness, we measured the ability
of jammed microphones to record human speech for recognition and content
extraction.  We tested the two jammers using built-in microphones of three different
smartphones: iPhone X (2018), Xiaomi Mi 6 (2017), and iPhone SE (2016).  We used the same experimental setup as
shown in Figure~\ref{fig:setup}.

In each experiment, we used a high-quality bluetooth speaker as a
proxy\footnote{We did experiments to study the potential difference between
  human speakers and bluetooth speakers by doing extensive recordings of
  both. For both audio spectrogram and speech recognition, the two are quite
  similar.} of a human speaker by playing pre-recorded human
speech at a standard sound level of human conversation (55-60dBA measured at
1m away according to~\cite{olsen1998average}). This setup avoided
inconsistency caused by potential participants and ensured a fair evaluation of the
jammers. The pre-recorded speech used in our experiments was taken from the LibriSpeech
dataset~\cite{librispeech}, which is commonly used by speech recognition
researchers, and includes randomly selected 1000 sentences of clean human speech.

We used two metrics to evaluate the jamming effectiveness. {\em First}, for
each test, we recorded audio on the target smartphone under active jamming, and
used the recordings to compute the Perceptual Evaluation of Speech Quality
(PESQ)~\cite{beerends2002perceptual}, which is an objective voice quality metric. The PESQ ranges between -0.5 and 4.5, where lower scores mean lower voice quality. {\em
  Second}, we fed our recordings into five popular speech recognition systems,
CMUSphinx~\cite{CMUSphinx}, Google Speech
Recognition~\cite{googlespeechrecognition}, Microsoft Bing Voice
Recognition~\cite{bingvoicerecognition}, IBM Speech to
Text~\cite{IBMspeechtotext}, and Kaldi toolkit with ASpIRE
model~\cite{povey2011kaldi}. The last two systems are particularly known for
their robustness against noisy speech signals. We picked the best speech
recognition results of these systems, and recorded its Word Error Rate (WER),
the common performance metric on speech recognition.  As a baseline,
we note that WER without jamming is around 30\% for the smartphones.

%\centering
%\begin{minipage}{0.32\textwidth}%
%	\centering
%	\includegraphics[width=1\textwidth]{figs_new/amz_effectiveness/amz.pdf}
%\vspace{-0.15in}
%	\caption{WER of the recorded speech when the i4 jammer is off,
 %         and when it points directly to the target microphone (1m
  %        away, $\alpha=0^\circ$). } %
%	\label{fig:jamming_effectiveness}
%	\vspace{-0.15in}
%\end{minipage}
%\hspace{0.1in}
\begin{figure}[t]
\centering
\begin{minipage}{0.32\textwidth}
   \includegraphics[width=0.99\textwidth]{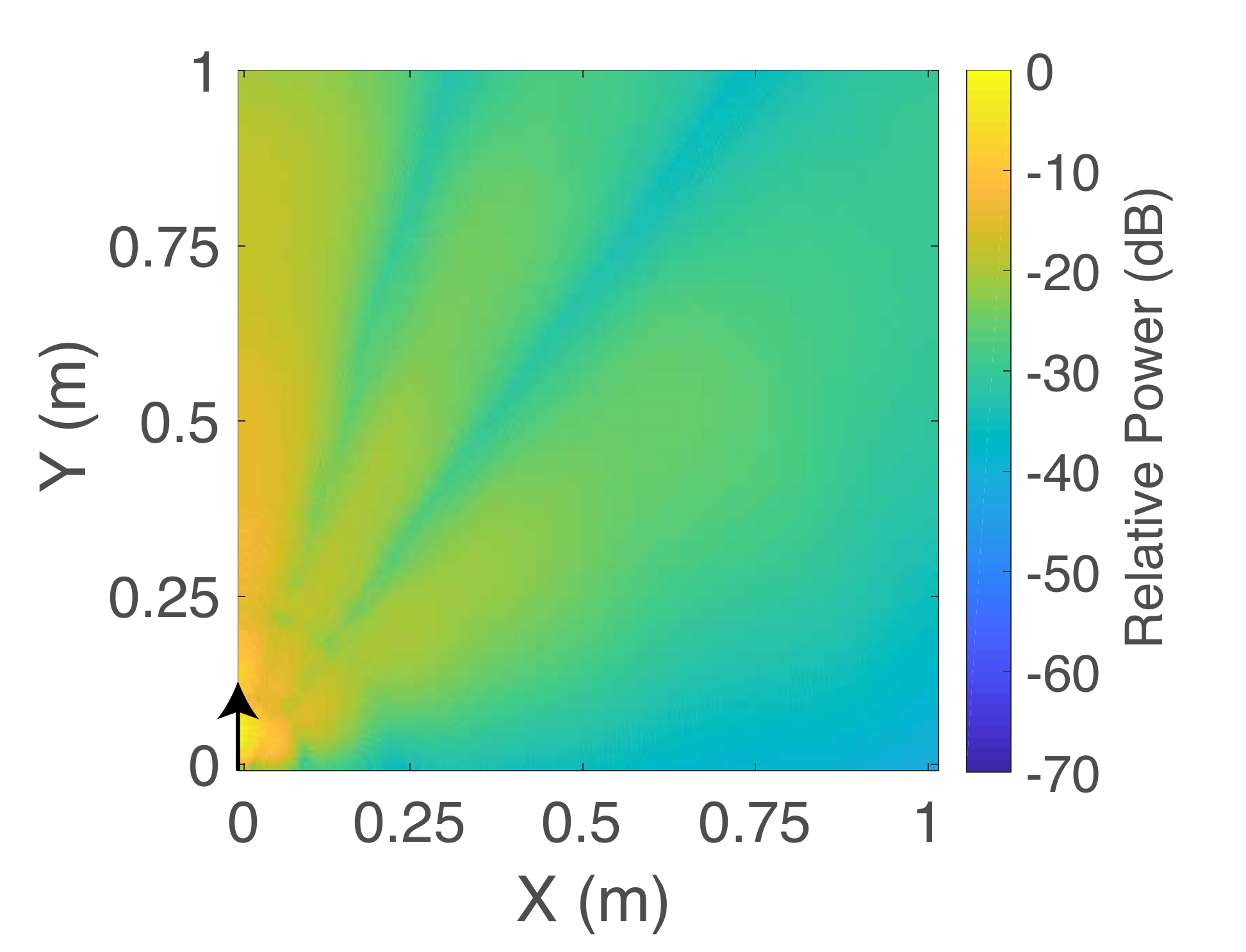}
\vspace{-0.2in}
   \caption{Simulated jamming power of the {\em Backdoor} 3$\times$3 jammer an 1m$\times$1m
     area. The jammer is placed at (0,0) and points to  (0,1) (marked
     by an arrow). The i4 jammer shows a similar pattern. }
    \label{fig:heatmap3x3}
\end{minipage}
\hfill
\begin{minipage}{0.64\textwidth}
\subfigure[WER]{
   \includegraphics[width=0.48\textwidth]{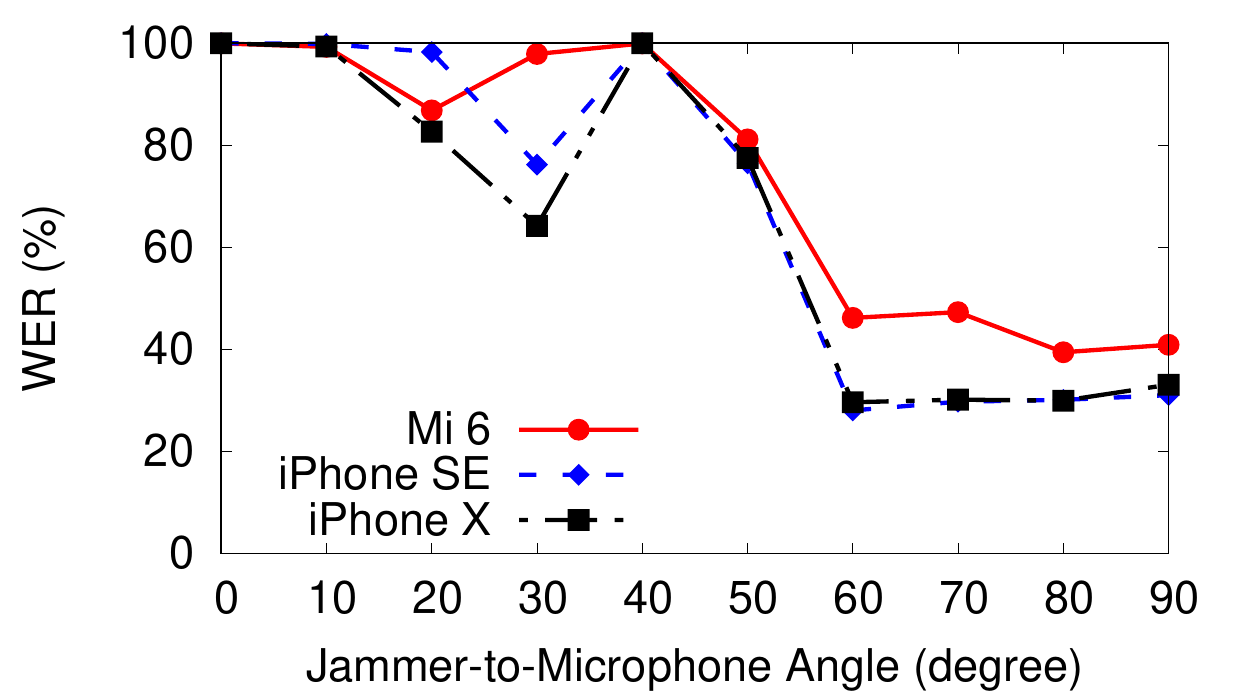}
   \label{fig:amz3}
 }
\hfill
\subfigure[PESQ]{
   \includegraphics[width=0.48\textwidth]{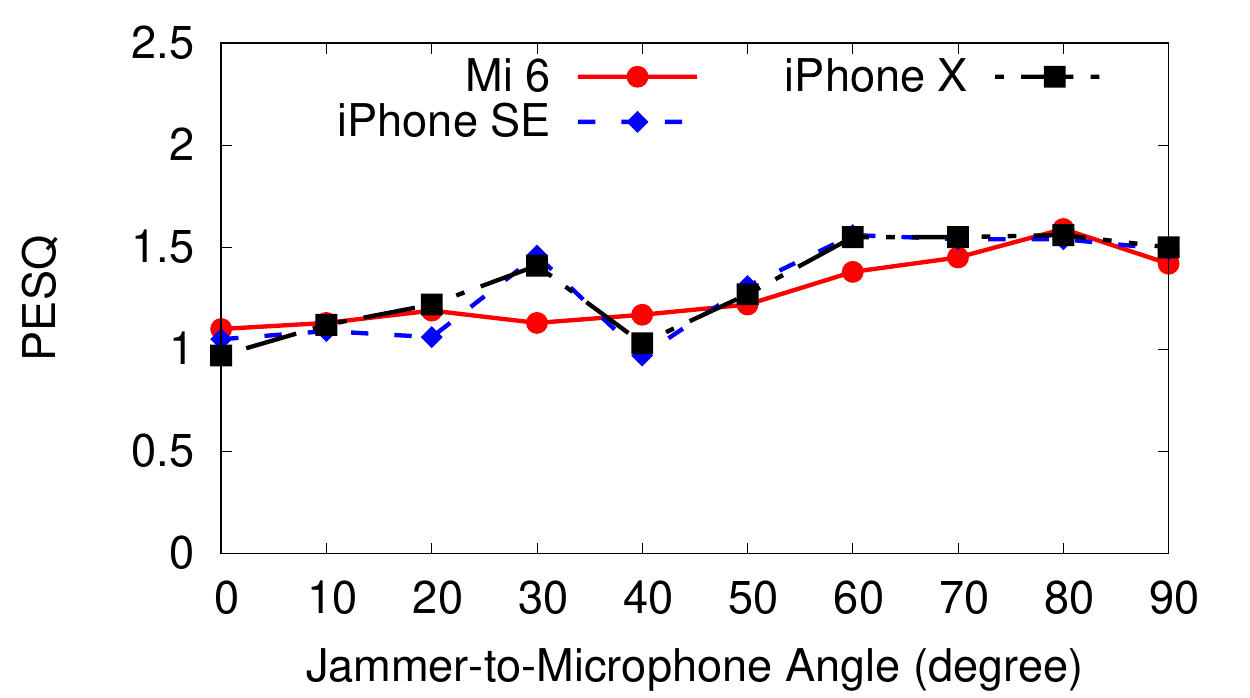}
    \label{fig:amz2}
   }
	\vspace{-0.15in}
\caption{WER and PESQ when the i4 jammer was 1m away from the target
  microphone at varying angular separation $\alpha$, using the same
  setup of Figure~\ref{fig:setup}. The {\em Backdoor} 3$\times$3 jammer showed
a similar trend (results omitted).}
\label{fig:eval_amz_angluar}
\end{minipage}
\end{figure}

% \para{Effectiveness of Directional Jamming.}  When the
% Figure~\ref{fig:jamming_effectiveness} plots the WER for all four
% microphones when the i4 jammer directly points to each of
% them ({\em i.e.\/} angular separation $\alpha=0^\circ$) and when the jammer is off.  We see that the jamming is quite
% effective, increasing WER from 20-30\% to nearly 100\%.  The same
% applies to the 3$\times$3 jammer and we omit the results for brevity.

%Also, as expected, the jamming effectiveenss reduces with the distance
%between the microphone and the jammer, since ultrasonic signals
%experience higher propagation loss than human voice signals~\cite{bass1984absorption}.

% We first look at the speech recognition performance when the jammer is pointing to the front side of the  microphones.
% As shown in Figure~\ref{fig:jamming_effectiveness}, for the selected smartphones, WER increases to 100\% when the jammer is on.
% Clearly, this means the ultrasonic jamming is highly effective.
% Also we found that the jamming effectiveness decreases with distance because the
% propagation loss of ultrasonic signals is higher than that of human voice
% signals~\cite{bass1984absorption}.

\para{Effectiveness of Omni-directional Jamming.}  For the i4 jammer,
Figure~\ref{fig:eval_amz_angluar} depicts the WER and PESQ results as a
function of the angular separation $\alpha$, for each of the phones
tested.  When pointing the microphone device (smartphone) directly
($\alpha=0^\circ$) at the jammer we observed a WER of almost 100\%.
But when angular separation $\alpha$ exceeded $50^\circ$, we observed a
significant drop in WER from 100\% down to 30-40\%, indicating that the jammer
was no longer effective.  PESQ also increased from 1 to beyond 1.5, following
the same trend (again, a larger PESQ means better voice quality).

Furthermore, the two iPhone models also showed large local fluctuations between
20$^\circ$ and 30$^\circ$, indicating the existence of blind
spots.  The effect was observed to a lesser extent on the Mi phone because its microphone is more
sensitive to jamming ({\em i.e.\/} by having a higher degree of non-linearity).  In the interest of visual clarity and brevity,
since the results of the {\em backdoor} 3$\times$3 jammer led to
similar observations, we omitted these.

\section{A Wearable Jammer Bracelet}
\label{sec:wearable}
Our evaluation results show that while ultrasonic signals can be used to
effectively disrupt microphone recordings, existing jammer designs offer very
limited angular coverage, and can only target microphones in a few specific
directions. This limits the privacy protection they might offer to users in
practice, since the user must know the location of nearby microphones and aim
their jammer accurately at the microphone.

In this paper, we consider a broader and more practical scenario where the
goal is to simultaneously jam and disrupt all microphones in an area around
the user. For this purpose, we explore the design of a wearable ultrasonic
jammer, which not only allows effective omni-directional jamming, but also
enables our solution to be highly portable, {\em i.e.\/} the jammer follows
the human speaker it is designed to protect.

In the following, we describe key design elements of our wearable jammer, and
describe how they overcome limited angular coverage and blind
spot issues faced by existing jammers (\S\ref{subsec:wearabledesign}). We
perform detailed benchmarks~(\S\ref{subsec:initialevaluation}), and
describe our current prototype as a self-contained bracelet
(\S\ref{subsec:prototype}).

\subsection{Key Design Elements}
\label{subsec:wearabledesign}

\para{Omni-directional Jamming via Circular Array.}  In theory, omni-directional signal
emission can be achieved using a circular array of ultrasonic
transducers, which emit signals simultaneously in many directions.  In
practice, the (angular) coverage of the jammer depends heavily on the number
of independent signal sources used to drive these ultrasonic transducers.

\begin{figure}[t]
  \centering
  \subfigure[w/ 24 independent sources]{
   \includegraphics[width=0.31\textwidth]{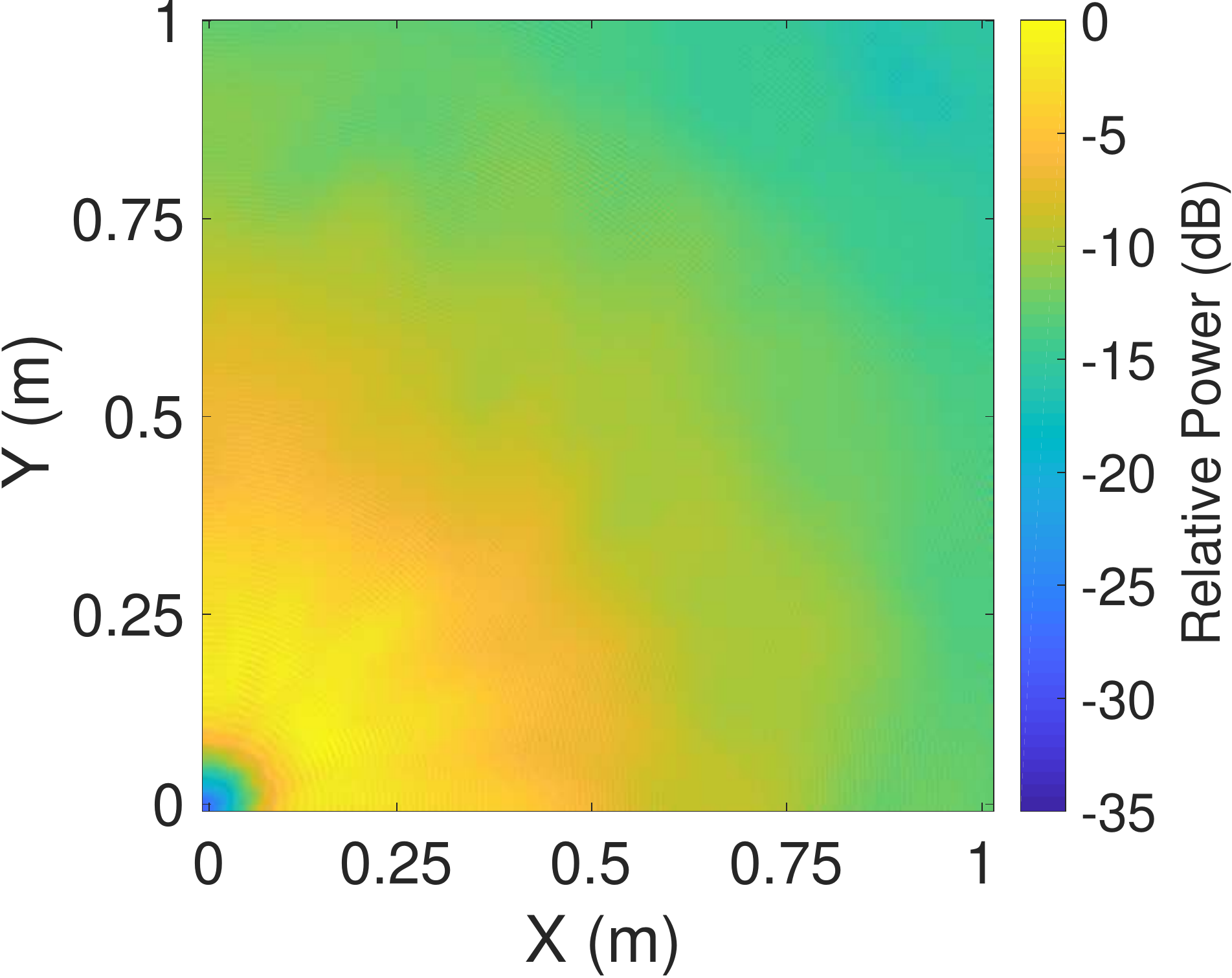}
 }
 \hfill
\subfigure[w/ 1 source]{
   \includegraphics[width=0.31\textwidth]{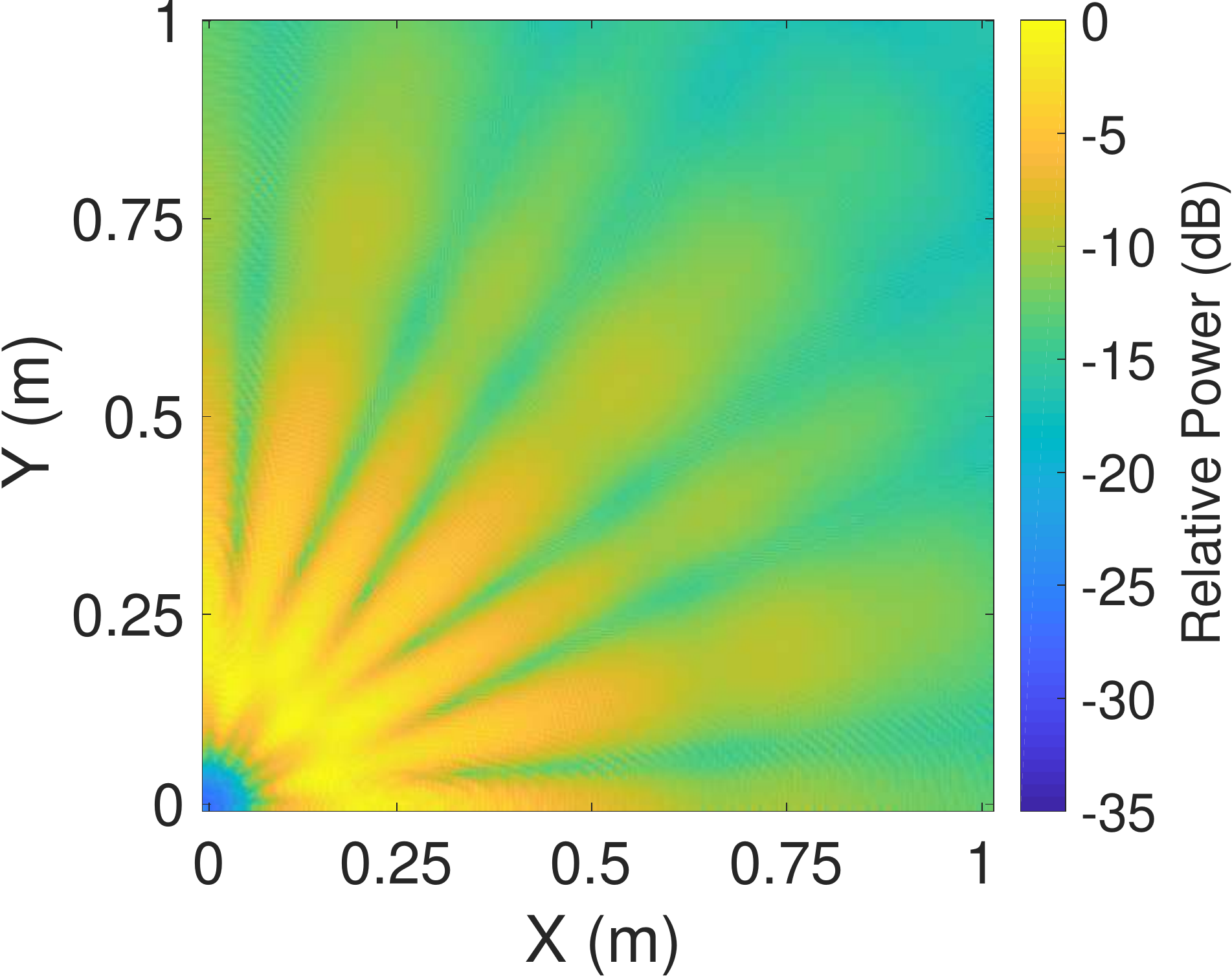}
   }
\hfill
\subfigure[w/ 2 independent sources]{
   \includegraphics[width=0.31\textwidth]{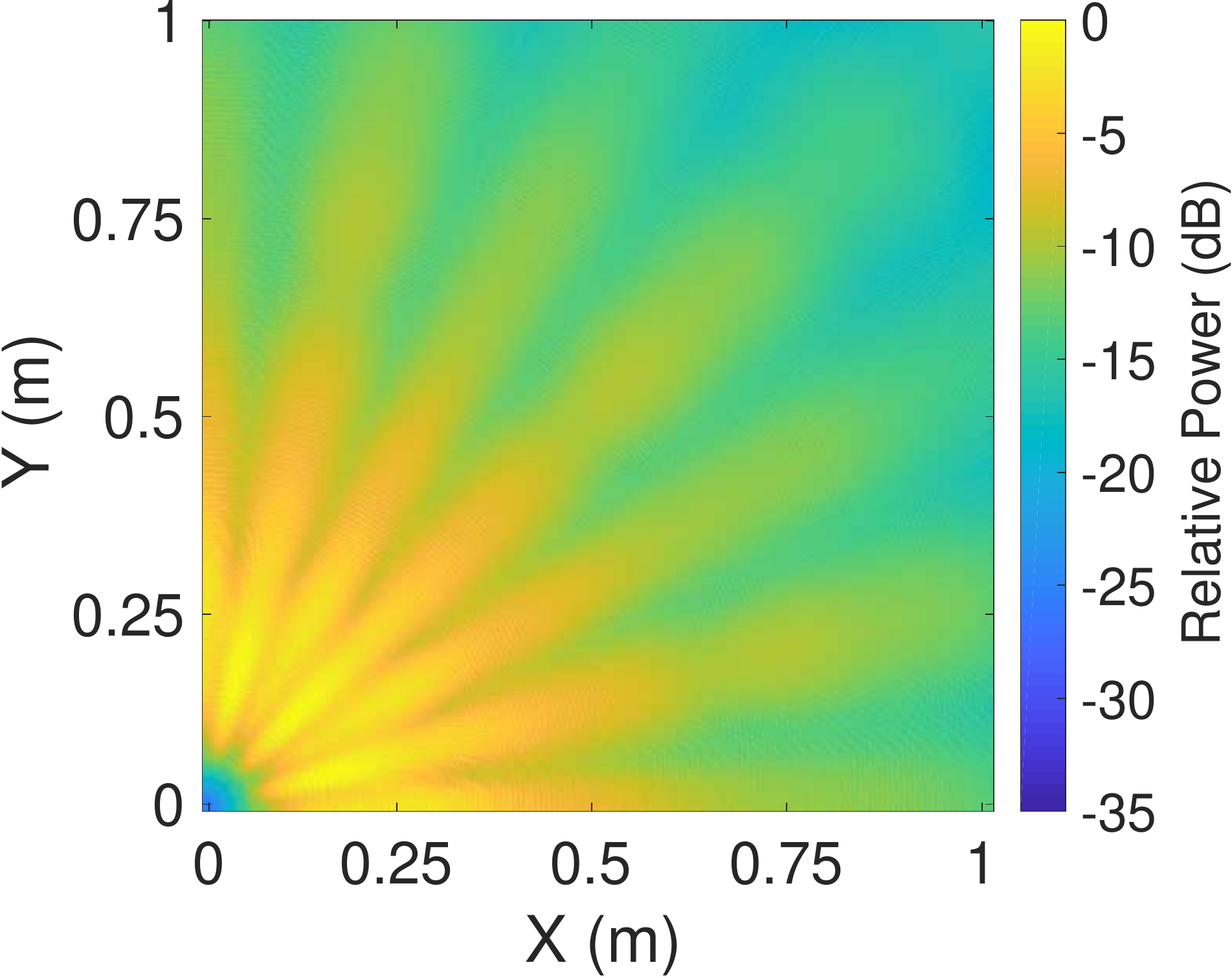}
   }

\vspace{-0.15in}
\caption{Simulated power map of a static wearable jammer in the form of a circular
array of 24 ultrasonic transducers, with 24, 2 and 1 input sources. The jammer
is placed at (0,0) in the 2D map but 10cm taller ({\em i.e.\/} on a user's wrist).  }
\label{fig:wearable_powermap}
\end{figure}

If the circular array can provide an independent input source for each
ultrasonic transducer, the simulated power map of the jammer will display a uniform
angular coverage, as in
Figure~\ref{fig:wearable_powermap}(a). But this is impractical for wearable
devices, since each input source is a high-resolution digital audio
player, and no more than 2 can fit on a form factor consistent with a single
wearable device.

When the array of transducers is driven by a smaller number of input sources,
interactions between transducers will again produce blind spots (or blind
angular directions), just like those produced by rectangular arrays in the
existing jammers.  Figures~\ref{fig:wearable_powermap}(b) and
\ref{fig:wearable_powermap}(c) plot the simulated power map where our
circular array jammer has one or two input sources.  While the jammer
radiates signals from all directions, we can observe multiple strips of locations where the jamming signal is 10dB lower
than nearby locations.

We note that for the above simulations,
the jammer is placed at (0,0) (as in Figure~\ref{fig:setup})  but 10cm taller
since it is now on the user's wrist (rather than the table).  As such,  the signal is weak at locations
within 5cm to the jammer due to the lack of vertical coverage.   This can be
addressed in practice by adding more transducers along the vertical
direction.

\para{Removing Blind Spots via User Movement and Gestures.}
A significant benefit to embedding the jammer as a wearable device is that we
can mitigate the blind spot problem by leveraging natural user movement,
% Interestingly, the above blind spot problem can be effectively
% addressed in practice as the human user moves naturally while
% speaking,
{\em e.g.\/} making a gesture or walking back and forth.  As the wearable
jammer moves with the user (her wrist), its instantaneous signal emission map
also changes.  Such natural signal fluctuations mimic the fading effect in
radio transmissions, creating instantaneous signal peaks and
valleys\footnote{Since the ultrasonic carrier frequency (25KHz) is much
  higher than that of human voice (85-180Hz), its signal fluctuation will be
  significantly larger and more frequent than that of human voice.}.  These
frequent signal peaks, although short in time, can effectively disrupt
the recording of individual words by a nearby microphone.

\begin{figure*}[h!]
\centering
\begin{minipage}{0.64\textwidth}
	\centering
	\includegraphics[width=0.99\textwidth]{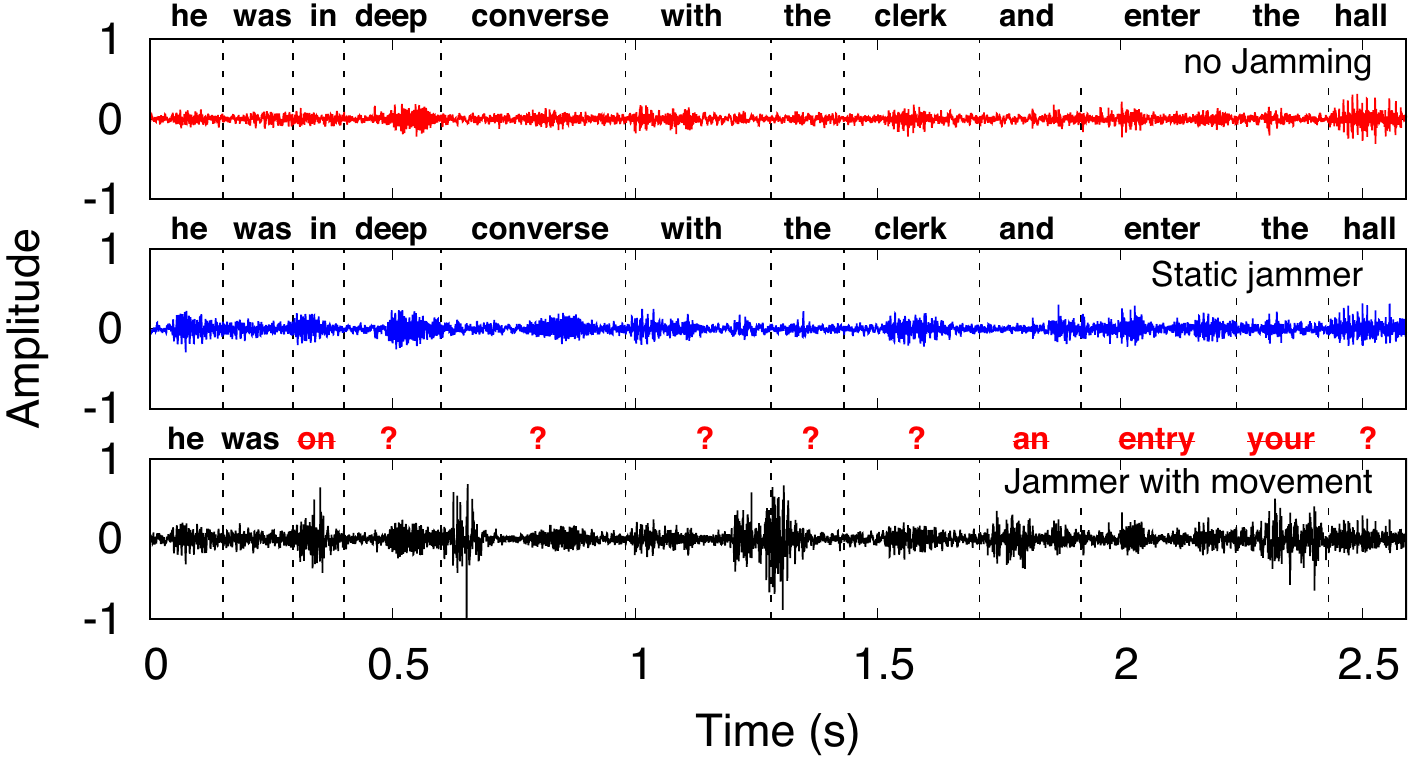}
        \vspace{-0.15in}
	\caption{A microphone is placed in the blind spot of a wearable
          jammer. If the jammer is static, its jamming has minimal impact on
          decoding human speech. The same jammer with small user gestures
          produces artifacts that make many words unrecognizable or
          introduce errors in word recognition.  }
	\label{fig:example_word}
      \end{minipage} \hfill
      \begin{minipage}{0.32\textwidth}
        \includegraphics[width=0.95\textwidth]{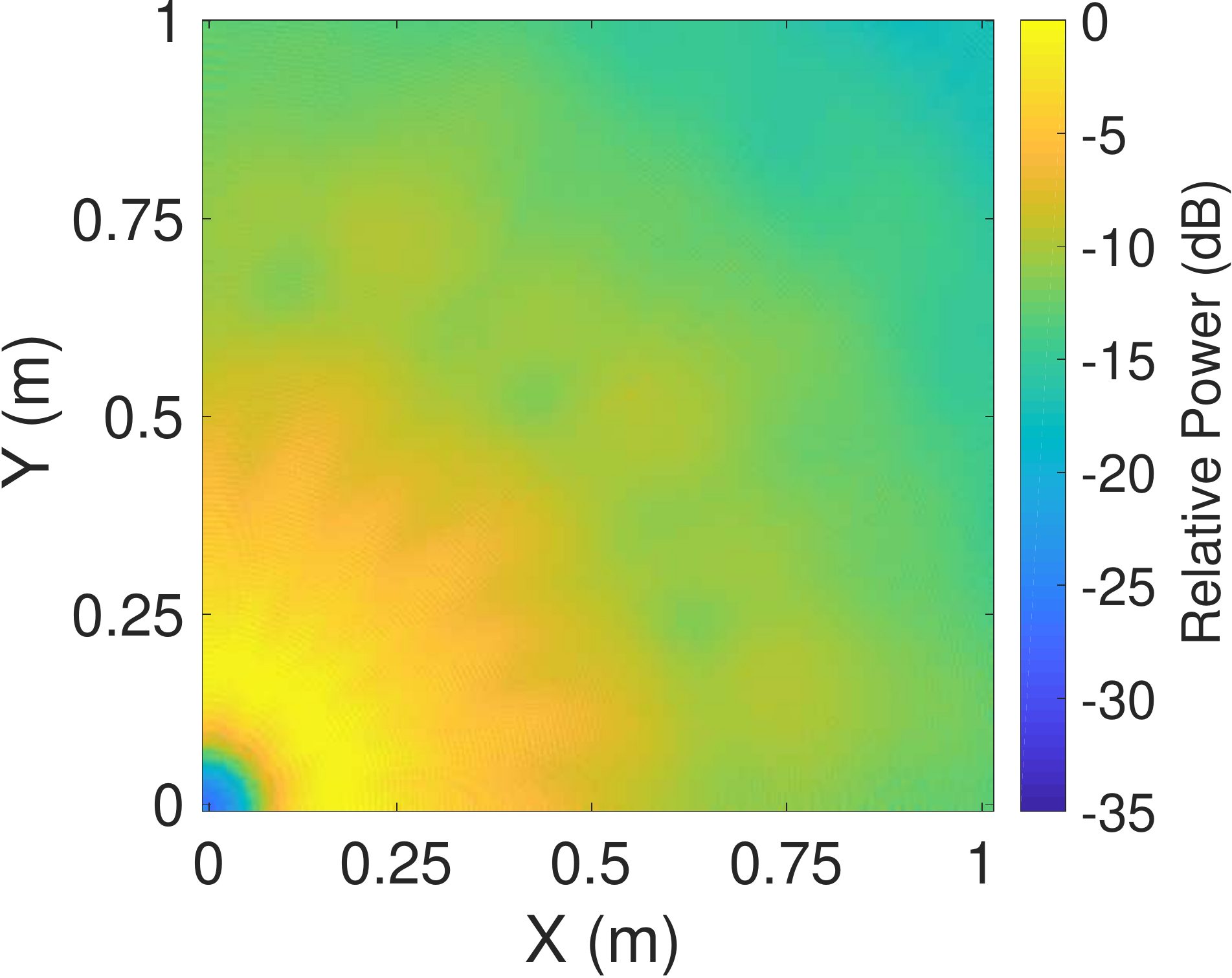}
              \vspace{-0.05in}
\caption{Simulated power map of a wearable jammer with 1 source  under small user
  gestures. The jammer is placed at (0,0) in the 2D map but 10cm
  taller ({\em i.e.\/} on a user’s wrist). } \label{fig:powermap_movement}
\end{minipage}
\end{figure*}

We test the impact of movement on jamming efficacy through an experiment. Using
a jammer with 24 transducers, we carefully place a microphone in its blind
spot. We then test the ability of the microphone to pick up and decode
pre-recorded human speech from the LibriSpeech dataset~\cite{librispeech} in
3 scenarios: normal recording (no jamming), static jammer (user staying
completely still), and jammer with movement (user with small
gestures). Decoded results in Figure~\ref{fig:example_word} show that a
microphone in a jammer's blind spot can decode human speech near perfectly,
but even small gestures are enough to make the majority (10 out of 12) of
words in the test unrecognizable. The two unaffected words were short,
monosyllabic words that recorded before jammer gestures started disrupting
the signal.

% an illustrative example in terms of the
% decoded speech signals and decoded words of a sentence, for the scenarios of
% no jamming (or jammer off), jammer staying completely static, and jammer
% moving as a result of human gestures.  Here the jammer has 12 transducers
% with 1 input source; the microphone is placed at a blind spot of the jammer
% when it is static. It shows that the static jammer's power (at its blind
% spot) is insufficient to overcome human speech.  But with natural user
% movements, the jammer makes 10 out of 12 words in the sentence
% unrecognizable.  The two unaffected words (``he'', ``was'') are very short,
% thus harder to break.
%But most of these short words carry much less weight in terms
 % of  security/privacy compared to longer words. \fixme{add citation?}

With this in mind, Figure~\ref{fig:powermap_movement} plots the
computed signal power map of the wearable jammer with natural human
movements (random rotation by up to 45$^\circ$), but with only a single
input source.  Since the
instantaneous jamming signal fluctuates significantly, we show the map averaged
over a window of 0.4s (average duration of a human spoken
word~\cite{spokenwordduration}). The resulting map closely approximates the
signal map of the (oracle) case where each transducer has its own input
source (shown in Figure~\ref{fig:wearable_powermap}).

% Our solution is to alleviate the weak spot effect is by moving the ultrasonic microphone jammer, {\em e.g.\/} a user is making a gesture or walking with the jammer while speaking.
% While the ultrasonic microphone jammer changes its location, the weak spots will also change their locations.
% Then as long as we keep moving the jammer, the microphone will not be always in the weak spots and we can guarantee the jamming effectiveness.
% Figure~\ref{fig:example_word} shows an example of ultrasonic microphone jamming when the user is speaking a sentence.
% We compare the cases of when the jammer is off, the user is static when the jammer is on and the user is making a gesture when the jammer is on.
% In the static case, the microphone happens to be placed in the blind spot of the jammer.
% From the figure, we can see that in both static and moving cases, the ultrasonic jamming is able to change the amplitude of the clean speech signal.
% But change is subtle for the static jamming case due to the weak spot effect, and speech recognition is still able to work.
% With moving, the change becomes significantly larger,
% and it's so large that the speech recognition completely fails for the whole sentence.
% This confirms that natural movement is able to reduce the weak spot effect.
% To get natural
% movement, we design the wearable jammer as a bracelet.

\para{Colocation with the Human Speaker.}  A wearable jammer is always
co-located with the human speaker it seeks to protect.  This not only
increases coverage (since the jammer moves with the user), but the short
distance between the jammer and the speaker's vocal cords also prevents
the use of beamforming microphone arrays to separate the signals of the human
speaker and the jammer~\cite{anguera2007acoustic}.
%heymann2016neural,xiao2016deep,li2016neural}.

%\begin{minipage}{0.32\textwidth}
%	\centering
%	\includegraphics[width=1\textwidth]{figs_new/simulations/gesture_rotation/quantile.pdf}
%\vspace{-0.15in}
%	\caption{Simulated power (over the space) of 7 degrees. A quantile figure}
%	\label{fig:multi_deg_rotation}
%	\vspace{-0.15in}
%\end{minipage}
%\hfill
%\begin{minipage}{0.32\textwidth}
%	\centering
%	\includegraphics[width=1\textwidth]{figs_new/gesture/experiment.pdf}
%\vspace{-0.15in}
%	\caption{Speech recognition results of different gesture.}
%	\label{fig:asr_gesture}
%	\vspace{-0.15in}
%\end{minipage}

\subsection{Validation via Benchmarking Experiments}
\label{subsec:initialevaluation}
Next, we use detailed benchmark experiments to evaluate our wearable jammer
design. Again we consider the scenario in Figure~\ref{fig:setup}, but replace
the jammer with our wearable prototype (placed on the human speaker's wrist,
10cm above the table). We examine both signal power distribution and speech
recognition accuracy.

\begin{figure*}[t]
\centering
\begin{minipage}{0.45\textwidth}
	\centering
	\includegraphics[width=1\textwidth]{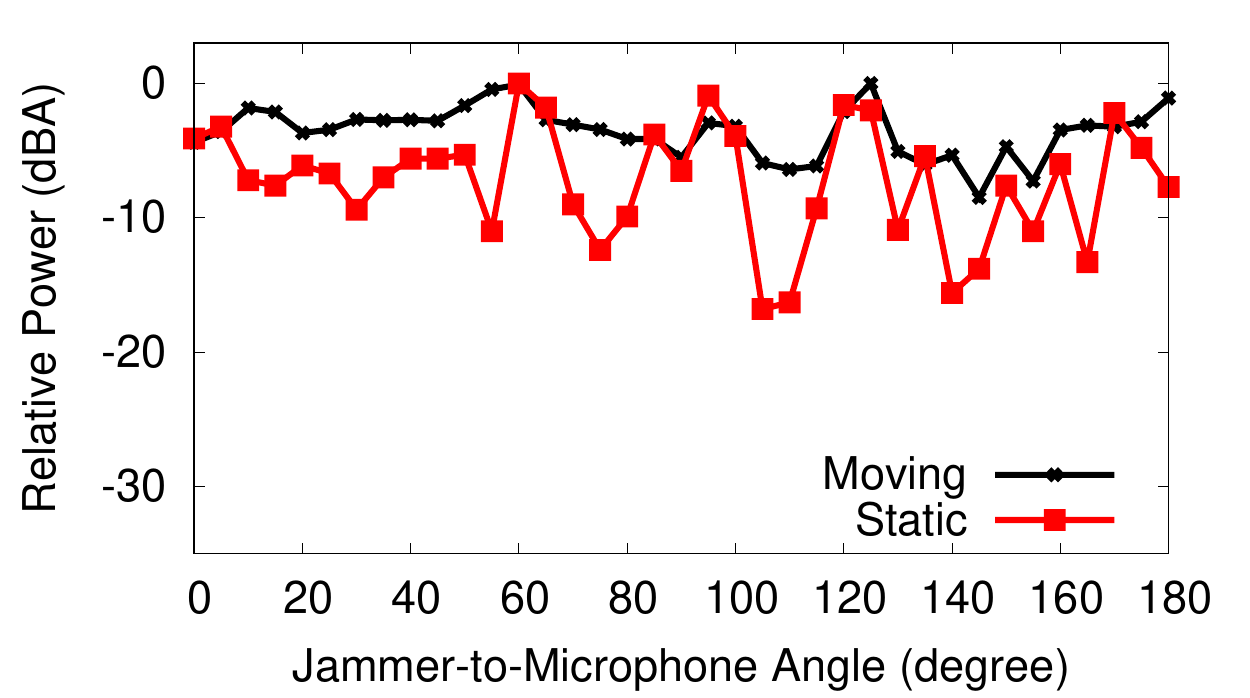}
\vspace{-0.15in}
	\caption{Angular coverage of the wearable jammer when
          completely static and under natural human movement. Jammer is 1m away from microphone.}
	\label{fig:wearable_angular}
	\vspace{-0.15in}
\end{minipage}
\hfill
\begin{minipage}{0.5\textwidth}
	\centering
	\includegraphics[width=0.85\textwidth]{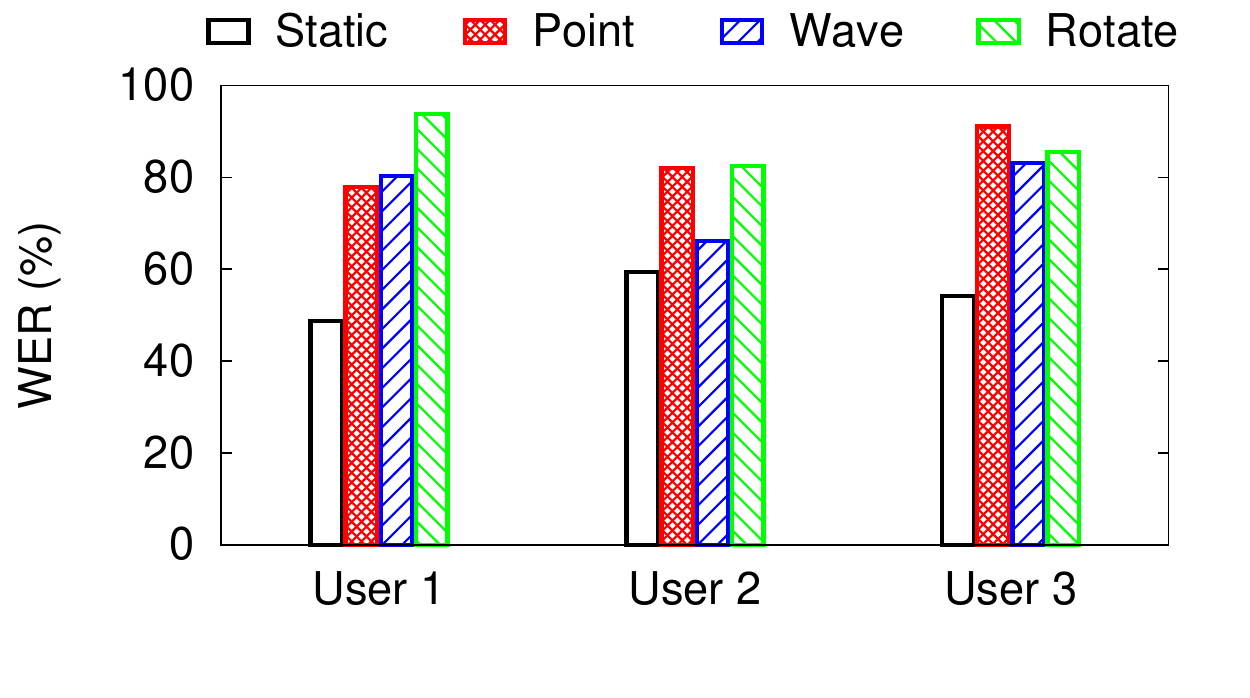}
\vspace{-0.15in}
\caption{WER of when each of the three human volunteers wears the
  jammer, and applies three gestures (point, wave, rotate) in their own
  styles. The microphone is placed at the blind spot of the jammer (when it
  is static).  The naturally-occurring human gestures largely increase jamming
  effectiveness. }
	\label{fig:asr_gesture}
	\vspace{-0.05in}
      \end{minipage}
      \end{figure*}

\para{Angular Coverage of Wearable Jammer.}
We first look at the angular coverage of our wearable jammer.   As in
Figure~\ref{fig:setup}, we measure signal power when the sound meter is
1m away from the jammer, but with different angular separation $\alpha$ to the
jammer/human speaker.  Figure~\ref{fig:wearable_angular} shows the
measured signal power (normalized by the highest power seen across
$\alpha=0^\circ$ and $180^\circ$), for cases where the jammer is completely
static and moving with the wearer.  Clearly, the jammer's movement helps to
smooth the signal power across $\alpha$, effectively removing blind
spots to offer omni-directional jamming.

\para{Speech Recognition Accuracy.}  We also validate the benefits of natural
jammer movement. In each experiment, while keeping the jammer static, we
identify a blind spot (1m away from the jammer) and place a microphone device
there.  We ask our volunteer to make one of the three popular gestures
(point, wave, rotate) as suggested by~\cite{gestures}, and run the
experiment for 30 sentences per gesture per round.  To minimize inconsistency in
human speech, we again use the bluetooth speaker to replay the same human
speech audio clip.

Figure~\ref{fig:asr_gesture}
shows the  WER for three volunteers.  We can see that naturally occurring hand
gestures greatly increase jamming effectiveness and remove blind spots. The
resulting WER increases to 70-92\%.

\subsection{Prototype}
\label{subsec:prototype}
We choose to design the wearable jammer as a bracelet that can be easily activated~\cite{Zeagler:2017iswc,Wang:2015ubicomp,DeRussis:2013ubicomp,Pakanen:2014,Angelini:2013ubicomp} whenever
the user decides to engage in a private conversation.  We now provide all the necessary technical details to implement our prototypes. To assist readers in replicating our prototypes, we provide all microcontroller code, signal parameters, circuits, and 3D files\footnote{Anonymized for review.}.
%This provides a source of ``always available input''~\cite{Saponas2009}, ensuring the user is the one in control.

\begin{figure}[h!]
    \centering
    \mbox{
        \subfigure[Our initial prototype of the jamming bracelet; in this prototype the signal generator, amplifier, and power supply sat outside the device (not shown).]{
            \includegraphics[height=8cm,keepaspectratio]{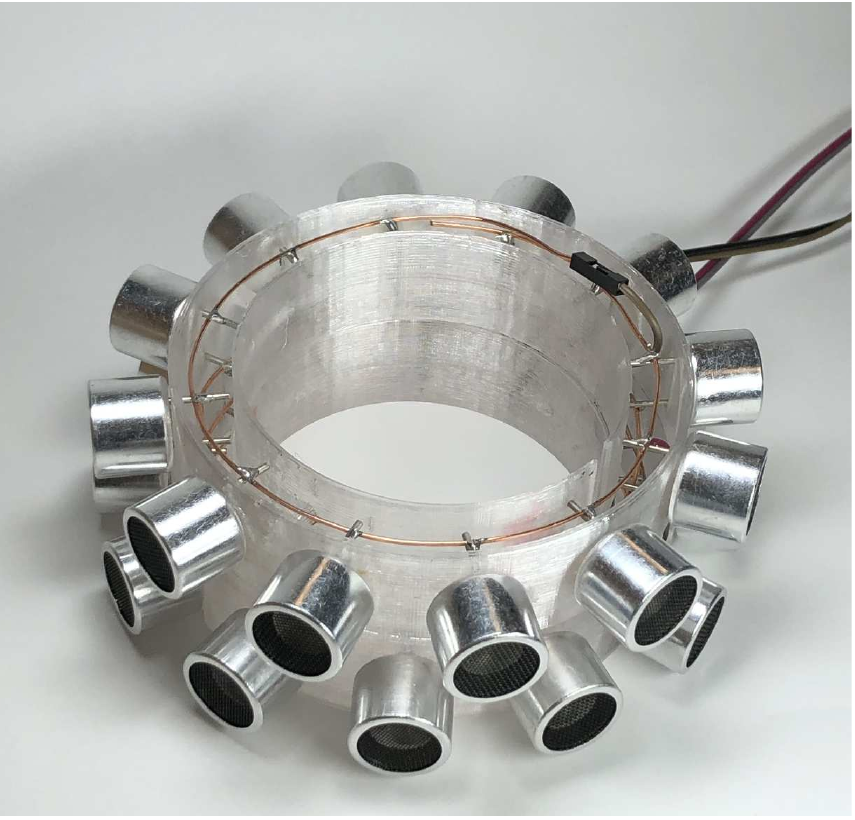}
        }
        %\hfill % doesn't work
        \hspace{0.4cm}
        \subfigure[Our final prototype is a self contained wearable device (battery, signal generator, microcontroller, touch button, LED status, and amplifier are all integrated).]{ 
            \includegraphics[height=8cm, keepaspectratio]{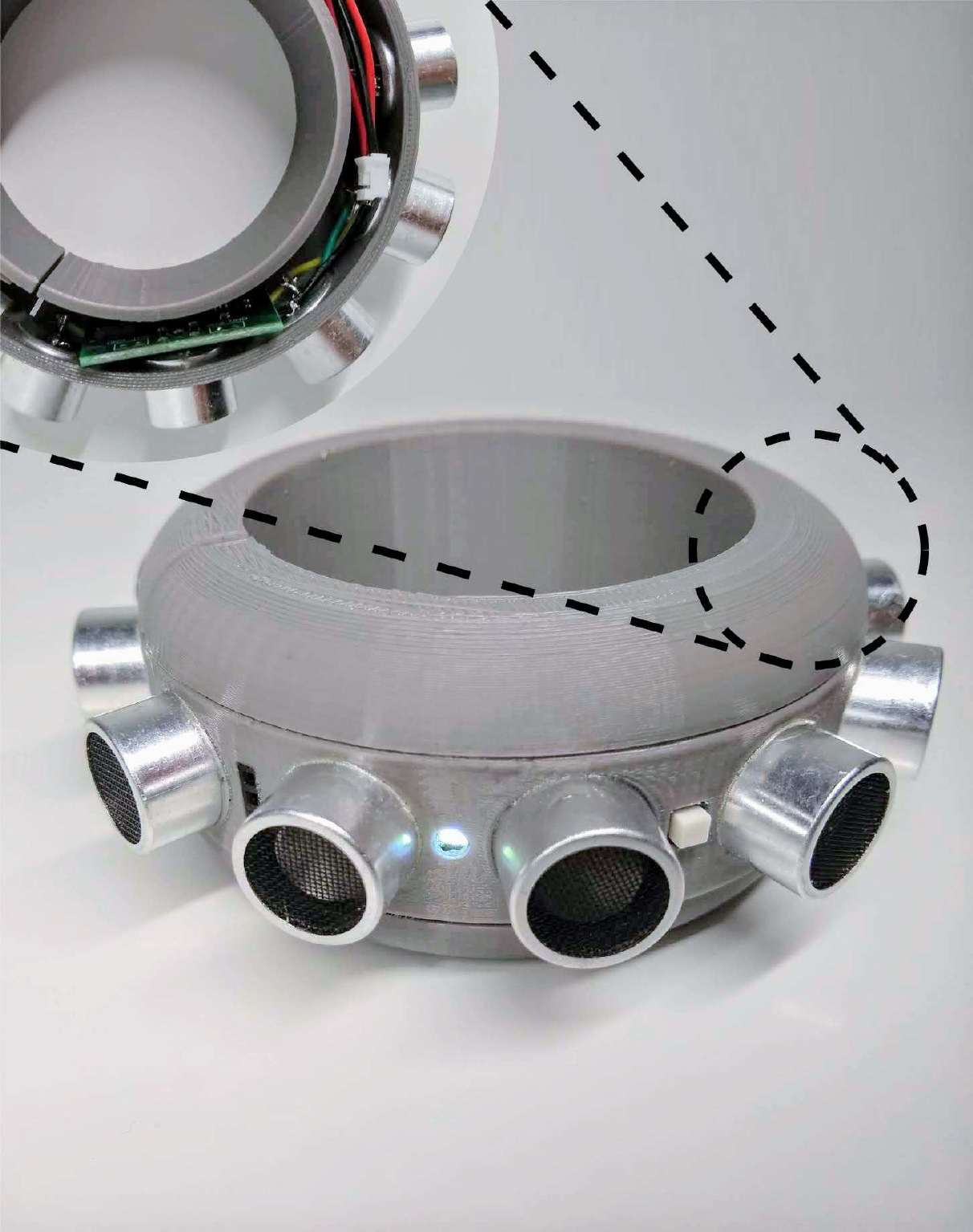}
        }
    }
    \caption{Our prototypes: (a) initial version used for experiments and (b) an improved and stand-alone wearable jammer.}
    \label{fig:bracelet}
\end{figure}

Our initial version, which is depicted in \ref{fig:bracelet}(a), of
our prototype bracelet is made from a simple 3D printed ring that
holds an array of 12 ultrasonic transducers (NU25C16T-1, 25kHz) evenly
spread in its perimeter.  These transducers all connect to a single
ultrasonic signal generator and an audio amplifier (PAM8403), which
both sat outside the bracelet. For simplicity in our initial version,
we used a Galaxy S7 edge smartphone as the signal generator, which is
capable of playing up to 192Khz through the line-out port. We
configure the signal generator to produce amplitude modulated white
noise centered on 25kHz ($\pm$1 Khz). We configured the audio
amplifier such that all the ultrasonic transducers operate at their
maximum power level. In all our experiments, we used two of these
bracelets stacked together (totalling 24 transducers), allowing us to
get a sense for the upper bound of the design. As our experiments
confirmed that making a jammer wearable does improve its effectiveness
({\em e.g.\/}, it reduces blind spots), we engineered an improved and stand-alone version.

Our improved prototype, which is depicted in
Figure~\ref{fig:bracelet}(b), is a self contained wearable device
comprised of the following components: a 3D printed shell, 12
ultrasound transducers (same as the above), a small low-powered signal
generator (AD9833, up to 12.5MHz with 0.004Hz programmable steps), a
ATMEGA32U4 microprocessor, an LED status, a touch button, and a small
rechargeable LiPo battery (105mA, which is 26 times smaller than an
iPhoneX's battery). The microprocessor controls the signal generator
via Serial Peripheral Interface (SPI). 

%To optimize for battery life, the microcontroller is set to
%low-power sleep mode for most of the cycles and wakes up every second
%to re-establish the signal generator output (for the next 1 second).

We measured energy consumption of our prototype bracelet. It consumes about 148mW when jamming (which is 28 times less energy than the i4 jammer). To put this into perspective, our jammer uses roughly {\em 15\%} the energy consumed by the internal WiFi module of a typical smartphone~\cite{Carroll2010,energywowmom16}. Our resulting device (including its battery) weighs 91 grams.

%  and our largest bracelet configuration
% with 48 transducers (and the amplifier) consumes 446 mW.

% \para{Current Limitations.} While we believe our prototype is a step towards a fully
% wearable jammer, it has limitations. First, we have not mounted the signal
% generator board onto the bracelet itself. This will require a smaller signal
% generator board, such as adapting the control unit of today's ultrasonic
% proximity sensors\footnote{The SparkFun Ultrasonic Proximity Sensor (HC-SR04)
%   includes a ultrasonic transducer, a receiver and a control unit. It is of size
%   45$\times$20$\times$15mm, unit weight 0.26oz, \$3.95, more info at
%   \url{https://www.mouser.com/datasheet/2/813/HCSR04-1022824.pdf}} to our
% system, but we do not anticipate this to offer any technical
% challenges. Second, we also did not include the on/off switch, which allows
% users to activate/deactivate the jamming, directly in the frame of the
% bracelet.

\para{Current Limitations.}  While we believe our prototypes are a step towards wearable jammers, they have limitations.
First, like all current ultrasonic jamming techniques, the user cannot selectively jam devices: i.e., a user cannot choose to avoid jamming their own smartphone while the signal is on. % OLD: "First, as we all current ulrasonic jamming techniques, the user cannot selectively choose to jam certain devices; i.e., it is not possible to not jam one's own smartphone while the signal is on."
On this limitation, our approach does provide much more control than existing stationary jammers, because the user does not have to walk all the way to the jammer to disable it and can do so by simply touching the bracelet. Furthermore, in Section~\ref{sec:discussion} we present some initial steps to further include selective jamming. Second, due to hardware limitations, the current prototype is larger than a typical bracelet and has limited vertical coverage.
However, we believe that switching to newly developed ultrasonic transducers -- like~\cite{CMUT} (which are 1.4 mm in diameter) -- enables the construction of a slim, stylish version of our wearable jammer. % OLD: "However, we believe that switching to newly developed ultrasonic transducers like~\cite{CMUT} (which are 1.4 mm in diameter) will allow to construct our wearable jammer as a slim and stylish bracelet."
Despite these shortcomings, we believe this prototype offers a great blueprint towards a low-cost and ubiquitous microphone jammer.

\begin{figure*}[t]
\centering
 \subfigure[One participant setup.]{
   \includegraphics[width=0.3\textwidth]{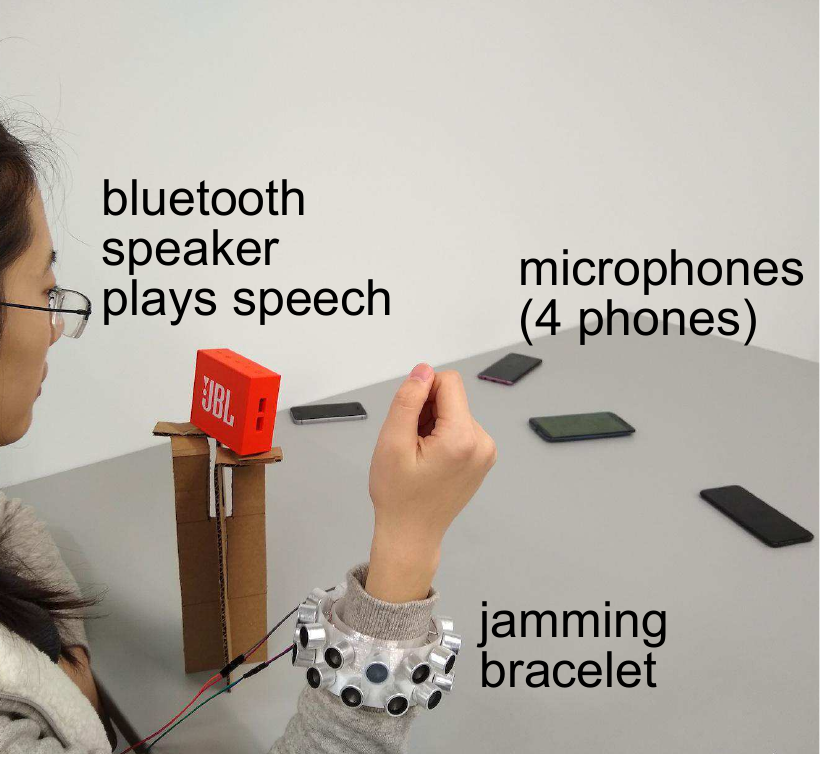}
   }
\hfill
%	 \subfigure[Hidden microphone example (under the t-shirt).]{
%   \includegraphics[width=0.40\textwidth]{figs_new/in_the_wild/setup/2.pdf}
%   }
%\hfill
	 \subfigure[Two participants around the microphones.]{
   \includegraphics[width=0.3\textwidth]{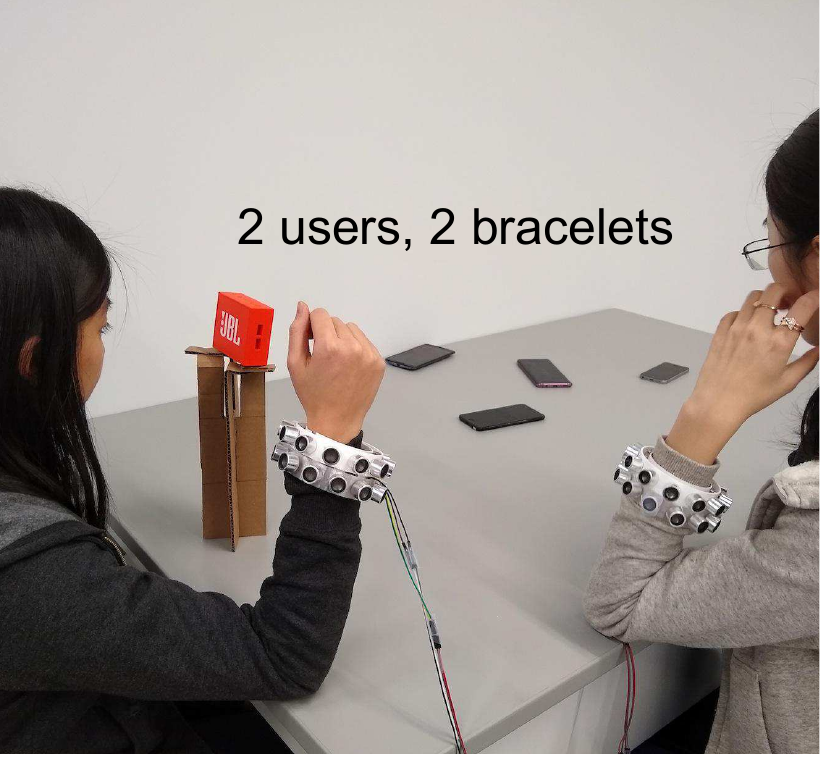}
   }\hfill 
   \subfigure[Participant walks back and forth in front of microphones.]{
   \includegraphics[width=0.3\textwidth]{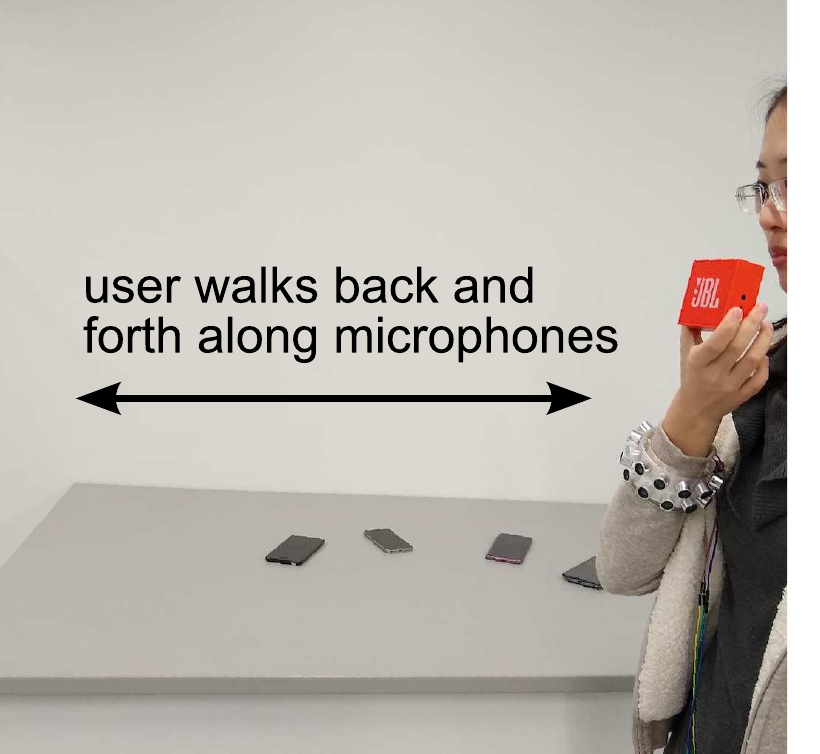}
   }
   \caption{User study setup we designed to investigate our jammer in more complex situations that a user might face at home, these include: walking around the room, interacting with other users, etc.}
	\label{fig:setups}
\end{figure*}

\section{Validation in Realistic Scenarios}
\label{sec:eval2}

In this section, we test the wearable jammer bracelet using four scenarios
designed to capture realistic situations that one might face at home or at
work, these are depicted in Figure~\ref{fig:setups}.

\para{Experimental Procedure.} We used a within-subjects design with 2 interface conditions (with our jamming bracelet or without) and 5 tasks. 

\para{Setup.} Participants were asked to wear our jamming bracelet (our
initial laser-cut prototype, featuring 2x12 transducers) on their dominant
arm. We left the participants in the experimental room for 80 minutes (as in
Figure~\ref{fig:setups}(a)), and they could do whatever normal daily
activities they wanted to, either sitting at or walking around the table, as
long as they completed the ``tasks'' we planned for them (see
below). Participants later reported activities such as looking at their
phones, reading books, etc. We also asked them to not speak, so we could
again use our bluetooth speaker playing pre-recorded speech as our consistent
proxy for a human voice.  Around the conference table, at a distance between
0.8m and 1m, we placed 4 different smartphones (Samsung Galaxy S9, Xiaomi Mi
6, iPhone SE, and Nexus 6), which we used to record the resulting jammed
speech that was later used to perform speech recognition.

\para{Tasks.} We asked participants to perform 5 tasks during their
experimental time: (1) sit on a chair at the table with a bluetooth speaker
positioned in front of them (see Figure~\ref{fig:setups}(a)); (2) same
activity, but with the bluetooth speaker positioned away from the participant to mimic another voice in the room; (3) work with another participant, where one was
tasked as the ``speaker'' while the other was a ``listener'' (see
Figure~\ref{fig:setups}(b)); (4) same as before but with the participants
flipping their roles as speaker and listener; and finally, (5) walk back and
forth in front of the table. Each task took 10 minutes.

\para{Participants} We recruited six participants from our local institution
for these experiments (aged 20-30 years old).

%Using this experimental setup we explored the effectiveness of our jammer bracelet from the following perspectives:

%\emph{first}, assuming the microphones are visible, such as a typical home-assistant device placed in the living room,
%\begin{packed_itemize}
%\item what is the effectiveness of the wearable jammer bracelet, and whether the bracelet can guarantee us secure conversation.
%\item if there are multiple people nearby, whether an extra person wearing the bracelet can help improve the jamming effect.
%\end{packed_itemize}

%\emph{second}, if the microphones are hidden, say, an attacker hides a microphone behind the curtain,
%\begin{packed_itemize}
%\item how much will the blockage reduce the jamming effect and how it varies from different blockage materials.
%\end{packed_itemize}

% \fixme{what is this doing here? This is a result.}
% We also observe that the PESQ and WER results as consistent:  the speech
% recognition system cannot recognize any speech segment with PESQ less than
% 1.5. When PESQ is above 1.5, speech recognition accuracy scales linearly
% with 
% PESQ.  For brevity, we will only present the WER result. 

\subsection{Jamming Visible Microphones}
We now detail the results for all tasks in which the microphones were placed
visibly on the table (as depicted in Figure~\ref{fig:setups}). These
placements are resemblant of how home-assistant devices are placed in a
user's living room. 

\begin{figure}[t]
\centering
 \subfigure[Single Speaker]{
   \includegraphics[width=0.44\textwidth]{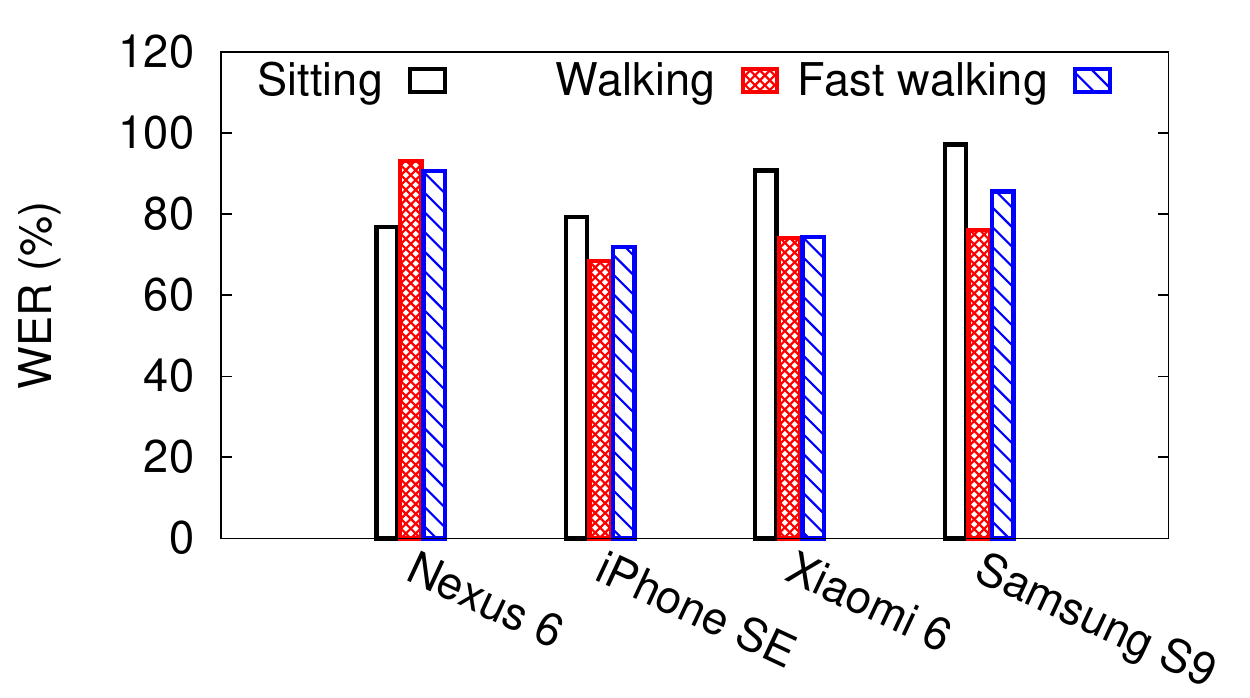}
   }
\hfill
	 \subfigure[Two-user Conversation]{
   \includegraphics[width=0.44\textwidth]{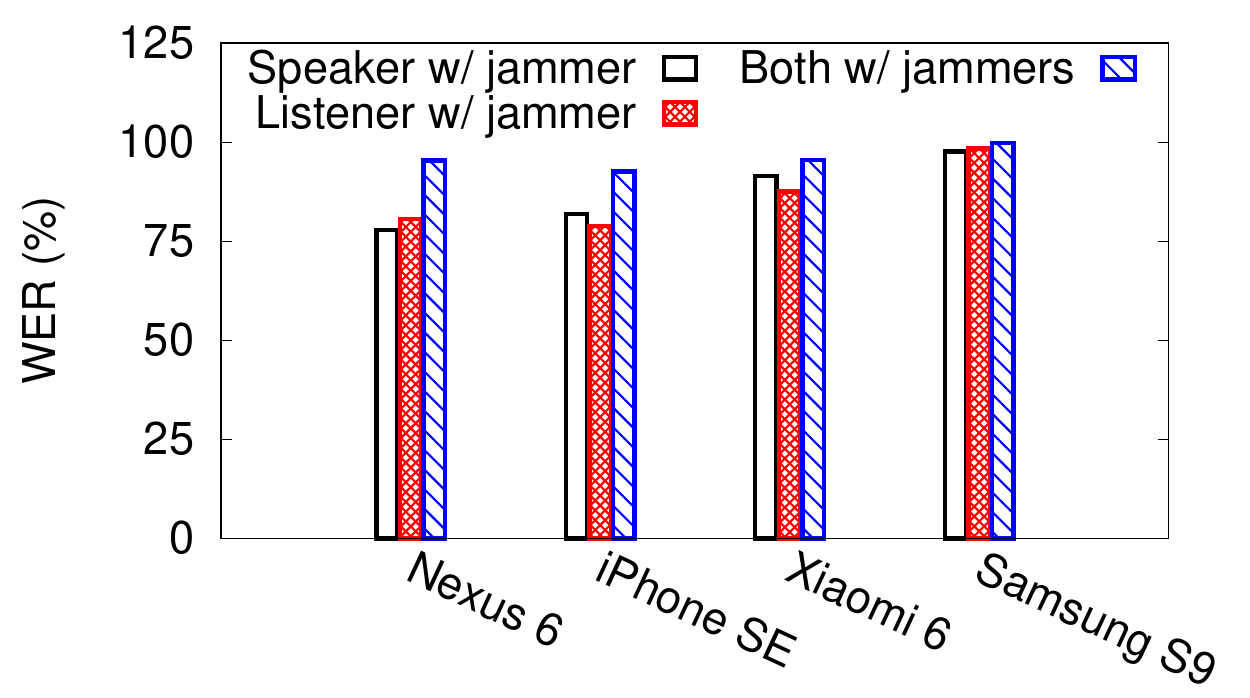}
   }
   \vspace{-0.1in}
   \caption{Speech recognition results for the tasks in which the microphones
   were placed visibily on the table. (a) A single participant, either sitting or
   walking. (b) Two participants, either one or both wearing our jammer.}
	\label{fig:in_the_wild}
\end{figure}

\para{Participant sitting down.} We compare the speech recognition results
when the participant is with and without the jammer bracelet. 
Results are quite consistent across participants, so we aggregate 
speech recognition results for all participants, shown in
Figure~\ref{fig:in_the_wild}(a).  Clearly, jamming was effective for
different smartphones positioned at different locations. Baseline WER was
30\% without our bracelets, and ranged between 75\% and 100\% when the
participant wore the bracelet.  The PESQ results are consistent and
thus omitted for brevity.

To see whether 75\% WER is sufficient to jam voice recordings, we look at the
recognized words.  Table~\ref{tab:wer_example} shows four examples of the
recognized sentence in different WER cases.  At a WER of 30\%, there was some
loss and mis-recognition. But at a WER of 75\%, the recognition results had
almost no overlap with the original sentence. 
% compared with the case of WER 30\%.  In the case of both simple and complex
% sentences, although our device jams 75\% there are some words that the
% recognizer is still able to detect. However, these 25\% remainder are not
% enough to help extract or infer reasonable meaning of the spoken sentences.
Further, we found that in higher WER cases, what few words were actually
recognized were not at all useful for understanding.  For example,
in WER of 99\% cases, we can only recognize words of ``and,'' ``do,'' ``sure,''
``his,'' ``show,'' ``this,'' ``make,'' ``to,'' ``of,'' ``for,'' ``the,'' ``think,'' ``is,''
``he''].

\para{Participant Walking Around.} We asked participants to randomly walk
near the table (distance within 0.8 m), while holding the bluetooth speaker
near their mouth in one hand, as shown in
Figure~\ref{fig:setups}(c). Participants were asked to walk in a mix of
styles, such as fast, slow,  and at their own pace. Each round takes
1 minute of walking.  Figure~\ref{fig:in_the_wild}(a) shows the speech
recognition results.  In all cases, WER with jamming is high ($>$70\%).
We observe no significant difference in speech recognition
performance when walking at different speeds.

\para{Multi-participant Scenario.} We consider a two-person scenario, to see
if another jammer-equipped person in the same room can help improve the
jamming effect.  We asked volunteers to work in pairs.  For each round of the
experiment, a pair of volunteers will sit in the room and do whatever they
want.  One will be the ``speaker,'' {\em i.e.} we place the bluetooth speaker
in front of them, and then we evaluate jamming effects where: 1) only the speaker
wears the bracelet; 2) only the listener wears the bracelet; 3) both
participants wear the bracelet.  Figure~\ref{fig:in_the_wild}(b) shows the
results. We see that if both of them wear the bracelets, the jamming performance is
the best - the WER can reach more than 90\% for all 4 microphones
(smartphones).  With an extra jammer, received jamming power 
increases and boosts WER.  When only one bracelet is worn, there is little
difference based on whether the jammer is worn by the speaker or the other
participant; both are able to disrupt the voice recording/recognition.

%Actually, the jamming effect depends on the distance and direction between
%the microphone (smartphones) and the speaker as well as the distance and
%direction between the microphone (smartphones) and the listener. 

\begin{table}[t]
\begin{tabular}{|l|l|l|}
\hline
\textbf{Sentence Transcript}                                                                                                                  & \textbf{Clean case of WER 30\%}                                                                                                               & \textbf{Jamming case of WER 75\%}                        \\ \hline
\small{Now to bed boy. }                                                                                                                              & \small{No too bed boy.}                                                                                                                          & \small{and they weren't too bad boy.}                     \\ \hline

\small{Gamewell to the rescue. }                                                                                                                              & \small{<noise> to the rescue.}                                                                                                                          & \small{You should be able to get rid.}                     \\ \hline

\begin{tabular}[c]{@{}l@{}}\small{Most of all, Robin thought of his father} \\ \small{what would he counsel. }\end{tabular}                                                                                                                              & \begin{tabular}[c]{@{}l@{}}\small{Most of all, Robin thought of his father} \\ \small{what would he counsel.}\end{tabular}                                                                                                                        & \small{List of a father we see counsel.}                     \\ \hline

\begin{tabular}[c]{@{}l@{}}\small{He began a confused complaint against} \\ \small{the wizard, who had vanished behind} \\ \small{the curtain on the left.}\end{tabular} & \begin{tabular}[c]{@{}l@{}}\small{You begin to confused complaint} \\ \small{against the wizard would vanished} \\ \small{behind the curtain on the left.}\end{tabular} & \begin{tabular}[c]{@{}l@{}}\small{Get confused complete to} \\ \small{get to the wizards.}\end{tabular} \\ \hline
\end{tabular}
\caption{\small{Examples of recognized sentence in clean speech case (WER 30\%) and jamming case (WER 75\%.)}}
\label{tab:wer_example}
\end{table}

\subsection{Jamming Hidden Microphones}
Next, we consider the task of jamming microphones hidden nearby, {\em e.g.} a
smartphone hidden inside a pocket or an attacker trying to stealthily record
a conversation by covering up a microphone.

\begin{figure}[t]
\begin{minipage}{0.5\textwidth}
\centering
\includegraphics[width=0.8\textwidth]{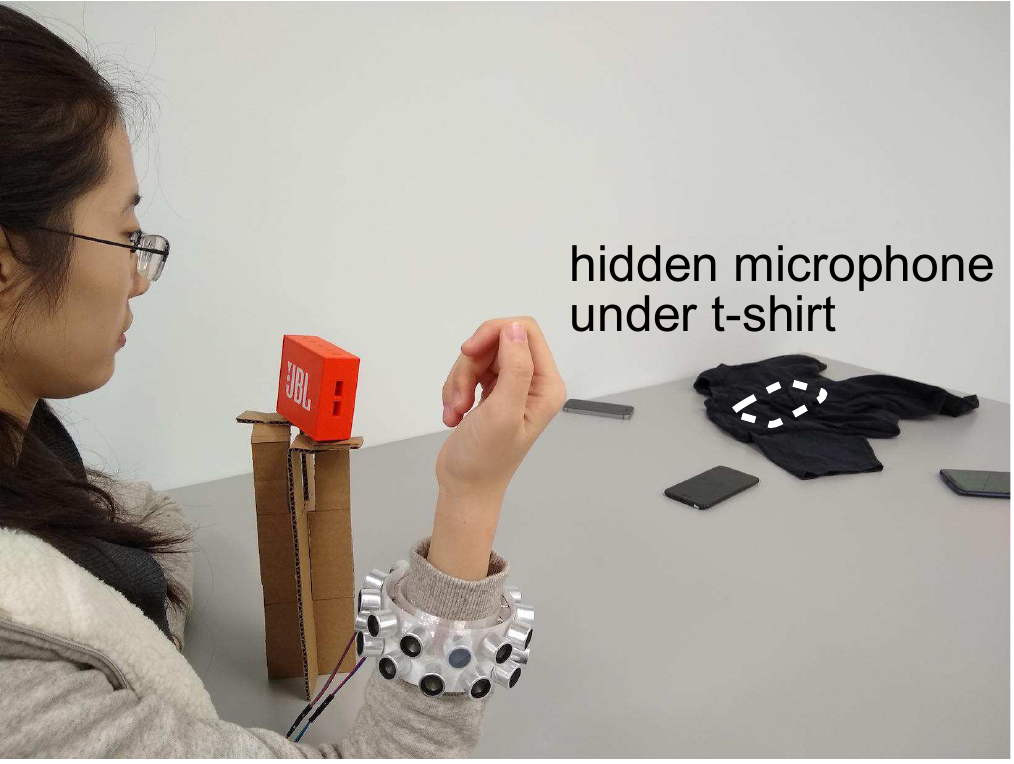}
\vspace{-0.15in}
   \caption{Experimental setup we designed to investigate how our jammer disrupts hidden microphones. Here, we depict one of the cases we explore, to hide a microphone under a T-shirt (but we also explored covering them up with boxes, etc).}
	\label{fig:hiddenmic}
\end{minipage}
\hfill 
\begin{minipage}{0.45\textwidth}
	\centering
	\includegraphics[width=1\textwidth]{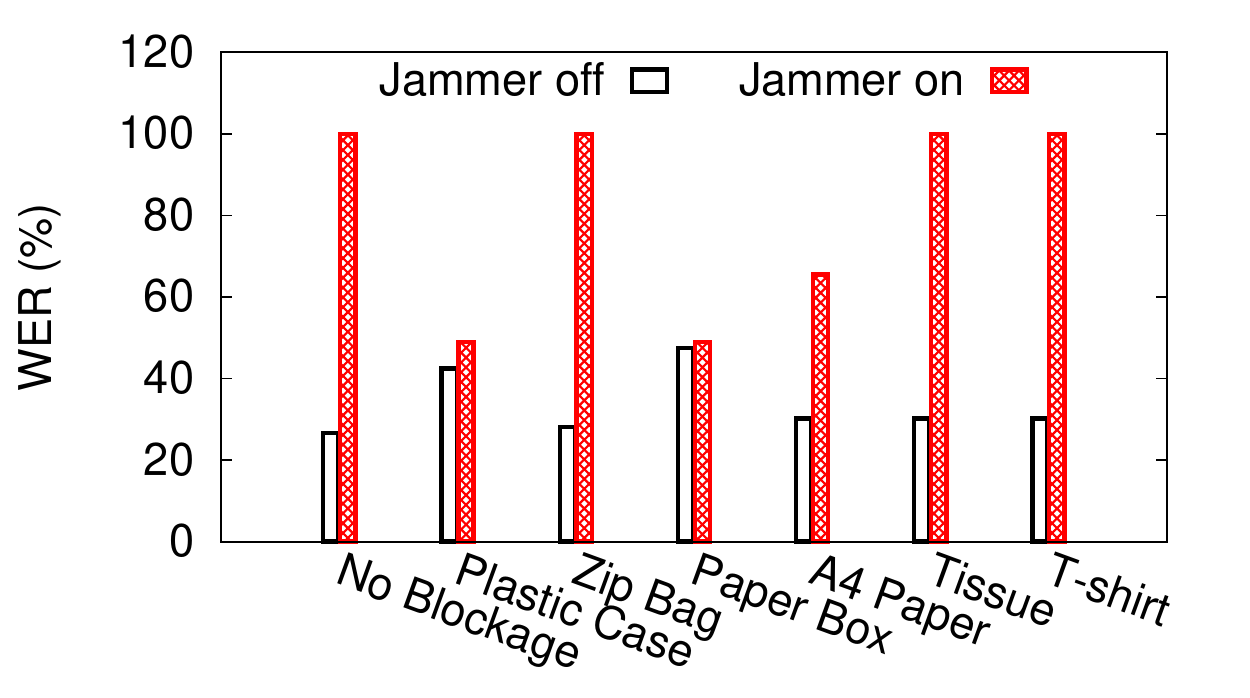}
\vspace{-0.15in}
	\caption{Speech recognition results when the microphone is covered up
        with various objects.}
	\label{fig:in_the_wild_blockage}
\end{minipage}
\end{figure}

To understand how different kinds of blockages can affect jamming
performance, we put different blockage materials on the microphone (shown in
Figure~\ref{fig:hiddenmic}). We then record the human voice audios with and
without jamming turned on.  From our results in
Figure~\ref{fig:in_the_wild_blockage}, we see that zip bags, tissues, and
T-shirts have little impact on jamming performance. Although the jamming
effect drops when blocked by A4 paper,  WER is still over 60\%. When the
microphone is covered by a plastic case or paper box, jamming performance
drops considerably, but the WER for clean audio also increases. In other words,
blocking our jammer signal also blocks normal audio. In general, most thin
blockage materials have little impact on jamming performance while
thick blockage materials will decrease the jamming performance. However,
thick blockage materials will also decrease the quality of the audio
recording.

\section{Discussion}
\label{sec:discussion}

% \subsection{Countermeasures by Eavesdropper}
% \para{Beamforming}
% \para{Ultrasound Microphone}
% \para{Extending 2D to 3D}

\subsection{Non-linearities of Microphone Hardware: Transient or Permanent?}

One may question whether the non-linearity of today's MEMS microphone
hardware is just a transient artifact of today's devices, and whether it will
disappear with improvement to microphone hardware. We believe non-linearity
is likely permanent for the foreseeable future, because MEMS microphones for
smartphones and voice-interface IoT devices are designed for low-cost and
small
form-factors~\cite{li-microphone17,lee-microphone08,boales-microphone17}.
Device manufacturers have little incentive to use materials with stronger
linearity properties, given the associated increases in cost and device size.
Our experiments on multiple smartphone devices released over the last 5 years show that non-linearity has not
decreased.

% and improvements in microphone sensitivity has actually made
%devices more susceptible to ultrasonic jamming.

\subsection{Ethics, Safety and Unintentional Jamming}
\label{sec:ethics}

Building mobile systems with ultrasonic signals requires an implicit
assumption that ultrasonic signals are harmless to surrounding users and
their devices. This assumption does not always hold.  There was even
speculation that high powered ultrasonic subharmonics have been weaponized to
produce undetectable discomfort to human targets~\cite{cuban-fu18}, though
the effects were later identified as being caused by the Indies short-tailed
cricket~\cite{crickets}. Here we discuss some considerations for ethics and
safety and possible risks of unintentional jamming.

\para{Risks and Experimental Precautions.}
Our proposed system uses ultrasonic frequencies in the 25kHz range, while the
upper limit frequency that the human ear can hear is between 15 and 20kHz.
The U.S. Occupational Safety and Health Administration (OSHA) warns that
audible subharmonics can be harmful at intense sound pressures of 105
decibels or above~\cite{ultrasafety}.
%We used a decibel meter to measure our
%sound pressure, which

To examine the safety of our jammer
device, we also measured the sound pressure level (SPL) of our prototype
bracelets using decibel meters.
Our prototype bracelets showed a maximum sound pressure of $<$100dB when measured directly at
the transducer, which quickly attenuates down to 73dB when 25cm away.

%\para{Sound Pressure Level (SPL).} Exposure to excess level of ultrasonic
%signals (155dB) can harm the human body by producing audible noise
%from source
%subharmonics\footnote{Research indicates while
 % ultrasonic signals have little effect on general health,  audible noise from
  %their source subharmonics can be harmful to human at very
  %high intensity~\cite{ultrasafety}.}.
%For this we used the RISEPRO Decibel Meter HT-80A\footnote{This
 % sound level meter has an accuracy level of +/-1.5dB and covers
  %frequency up to 40kHz.} and placed it near
%he bracelet.

% For our largest  bracelet of 48 transducers (or 2 bracelets of 24
% transducers each), SPL values at 0cm, 5cm, 10cm, and 25cm away from the
% bracelets are 100dB, 87dB, 85dB and 73dB, respectively, attenuating quickly in
% the air.  All of them are
% below the safety limit for airborne ultrasonic signals (105dB)~\cite{ultrasafety}.  Furthermore, while conducting experiments
% with our prototype bracelets (12 to 48 ultrasonic transducers at 25kHz) for many
% hours at a time, we experienced no pain or discomfort. We also did not
% detect any audible
% noise from the jammer.

During our experiments, coauthors conducted tests with multiple
bracelets (24 transmitting transducers),  for hours at a time, and reported no pain or
discomfort. All experiments were designed to keep jammers at least 30cm
away from human users' ears.

\para{Unintentional Jamming.}
\label{eval:phonecall}
It is possible that our jammer could accidentally jam legitimate microphones
in nearby IoT devices, including hearing aids or personal emergency response
devices. This is of course non-ideal. But given that our blocking range is
limited, we believe it would be easy for a user to detect an
unintentionally jammed person or device and turn off jamming as appropriate. More work is
necessary to understand the impact of ultrasonic signals on these devices and
to design workarounds.

Finally, we consider a scenario in which a user is trying to have a private phone
conversation with the jammer turned on. Users would like to avoid
inadvertently jamming their own phone. We have conducted some initial tests,
which show that, when a user speaks into her phone (held up to her face), she
can effectively block the ultrasonic jamming signal by shielding her phone
microphone and mouth with a hand.  Here the user's hand selectively blocks
the bracelet, but not her own mouth. As ongoing work, we are investigating
device designs that would make it easier for the jammer to speak into her own
phone.
% In this case, the smartphone's microphone
% used to make the call could be disabled by our jamming bracelets. To address
% this, we found that the user can use her own body to protect her phone's
% microphone from our jammer, {\em e.g.\/} placing the jamming bracelet on the
% left arm and using the right arm to hold the phone (to make the call) and
% turn her face to the right arm to block the microphone from the
% jammer. \fixme{Move entire para to discussion}

\section{Conclusions}
\label{sec:conclusion}

This paper describes our efforts in designing and validating a wearable
device that restores a user's sense of agency and control over their personal
voice privacy. Users can tap the device on, and be confident that their
conversations remain private from nearby microphones (both visible and
hidden).  As a disconnected device with no built-in microphone, our jammer
provides a tamper-resistant tool users can trust. Our prototype is
light-weight (91g), low-cost (\$36 for a complete 24-transducer bracelet,
including circuitry and battery), power-efficient (148mW), uses only commodity
components, and can effectively jam a range of listening devices from digital
assistants to smartphones.

There is certainly room for improvement in our current prototype. For
instance, recent developments in transducer design might result in
commercially available 1.4 mm diameter~\cite{CMUT} transducers; these can
dramatically reduce the bracelet size. This will open up new wearable form
factors that go beyond our bracelet, such as badges, rings and so forth. In addition, further tuning of the
ultrasonic signal will enable jamming over longer distances, while enabling
selective exceptions to more easily allow the user to carry on a phone
conversation.

%\section*{ACKNOWLEDGEMENT}
%We thank our anonymous shepherd and reviewers for their constructive
%feedback. This work is supported in part by NSF grants xxx.

{
 %\footnotesize
 \bibliographystyle{ACM-Reference-Format}
 \bibliography{ultra}
}

%\theendnotes

%\bibliographystyle{IEEEtran}
%\balance
%\begin{small}
%\balance
%\bibliography{zhao,translearn}
%\end{small}

\end{document}